\documentclass[twocolumn,pra,preprintnumbers,amsmath,amssymb,superscriptaddress,floatfix]{revtex4}%%
\usepackage{latexsym}
\usepackage{epsfig}
\usepackage{graphicx}
\usepackage{textcomp}
\usepackage{color}
\usepackage{mathtools}
\usepackage{lipsum}

\begin{document}

\title{Phase tracking for sub-shot-noise-limited receivers}

\author{M. T. DiMario}
\affiliation{Center for Quantum Information and Control, Department of Physics and Astronomy, University of New Mexico, Albuquerque, New Mexico 87131}

\author{F. E. Becerra}
\affiliation{Center for Quantum Information and Control, Department of Physics and Astronomy, University of New Mexico, Albuquerque, New Mexico 87131}
\email{fbecerra@unm.edu}

\begin{abstract}
Non-conventional receivers for phase-coherent states based on non-Gaussian measurements such as photon counting surpass the sensitivity limits of shot-noise-limited coherent receivers, the quantum noise limit (QNL). These non-Gaussian receivers can have a significant impact in future coherent communication technologies. However, random phase changes in realistic communication channels, such as optical fibers, present serious challenges for extracting the information encoded in coherent states. While there are methods for correcting random phase noise with conventional heterodyne detection, phase tracking for non-Gaussian receivers surpassing the QNL is still an open problem. Here we demonstrate phase tracking for non-Gaussian receivers to correct for time-varying phase noise while allowing for decoding beyond the QNL. The phase-tracking method performs real-time parameter estimation and correction of phase drifts using the data from the non-Gaussian discrimination measurement, without relying on phase reference pilot fields. This method enables non-Gaussian receivers to achieve higher sensitivities and rates of information transfer than ideal coherent receivers in realistic channels with time-varying phase noise. This demonstration makes sub-QNL receivers a more robust, feasible, and practical quantum technology for classical and quantum communications.
\end{abstract}

\maketitle

\section{Introduction}

Optical communication with coherent states can achieve the highest rate of information transfer through lossy and noisy channels \cite{giovannetti04, giovannetti11, mari14}. Coherent optical communications encode information in the coherent properties of the electromagnetic field, allowing for using high-spectral efficiency modulation and high-sensitivity coherent detection \cite{winzer12, kikuchi16}. Efficient coherent modulation and detection can dramatically increase the rate of information transfer beyond the reaches of intensity encodings \cite{ip08, li09, kikuchi16}. Moreover, the intrinsic nonorthogonality of coherent states can enable quantum communications \cite{arrazola14, clarke12, xu15} including quantum key distribution \cite{bennett92, huttner95, grosshans03, takeoka14, pirandola17, ghorai19} for secure communications over optical networks \cite{tang16, sasaki11}. However, coherent encodings are highly susceptible to phase noise and random phase variations in real-world devices and communication channels \cite{ip08, kikuchi16}. To ensure the expected advantage of coherent communications over intensity modulation and direct detection, communication protocols require efficient methods for phase estimation and phase tracking to correct for random phase changes induced by the channel \cite{ip08, li09, kikuchi16}, while being compatible with existing communication technologies. Moreover, practical scenarios in low-power and quantum communications require phase tracking based only on the transmitted signal state, without relying on transmissions of strong pilot phase reference pulses \cite{armada98, qi07, jouget13, qi15, soh15, huang15, marie17}.

Conventional coherent receivers that realize Gaussian measurements, such as heterodyne receivers, can perform phase tracking based on signal post-processing in the digital domain with diverse and efficient methods for channel and phase estimation \cite{barry92, lygagnon06, ip07, morsyosman11,wang19}. These methods renewed interest in coherent communications for increasing information transfer, and has made coherent communications more practical for future realizations of high-capacity communication networks \cite{he14, kikuchi16, gisin16}.

Further developments in optical communication will seek to approach the ultimate limits of information transfer in realistic communication channels. Quantum information science (QIS) provides the basis for approaching the fundamental limits in receiver sensitivities \cite{helstrom76} and information transfer in communications \cite{giovannetti04, giovannetti11, mari14}. {\color{black} Receivers based on Gaussian measurements, Gaussian operations, and local operations and classical communication have been investigated for information processing, phase estimation, and state discrimination \cite{weedbrook12}. The optimal Gaussian receiver for the discrimination of two nonorthogonal coherent states is the simple homodyne receiver \cite{wittmann10}. Furthermore, measurements based on adaptive homodyne detection can provide advantages for single-shot phase estimation of coherent states \cite{wiseman95,wiseman98,giacomo96}. However, the ultimate limits of receivers based on Gaussian operations for state discrimination are still under investigation \cite{weedbrook12, chesi18}}.
Among technologies enabled by QIS, non-conventional receivers, termed quantum receivers, use optimized non-Gaussian measurements based on photon counting \cite{kennedy72, dolinar73, bondurant93, cook07, wittman08, wittmann10, wittmann10b, tsujino11, becerra11, muller12, becerra13, izumi13, nair14, muller15, becerra15, bina16, ferdinand17, dimario18,dimario18b, dimario19} to provide sensitivities surpassing quantum noise limit (QNL) of coherent receivers \cite{kikuchi16}, and approach the true quantum-mechanical limit, the Helstrom bound \cite{helstrom76}.

Moreover, non-Gaussian receivers performing joint measurements over coherent-state codewords hold promise to bridge the gap between the Shannon and the Holevo limits in capacity \cite{giovannetti04, guha11}. However, making non-Gaussian receivers practical for coherent communications in realistic channels will require novel approaches for performing efficient phase tracking. These approaches will be fundamentally different from those based on conventional heterodyne detection using digital signal processing post-measurement \cite{barry92, lygagnon06, ip07, morsyosman11,wang19}, and will require realizing active phase estimation \cite{bina16, izumi16} and correction in real time, while ensuring performance beyond the QNL.

\begin{figure*}[!tp]
	\centering
	\includegraphics[width =\textwidth]{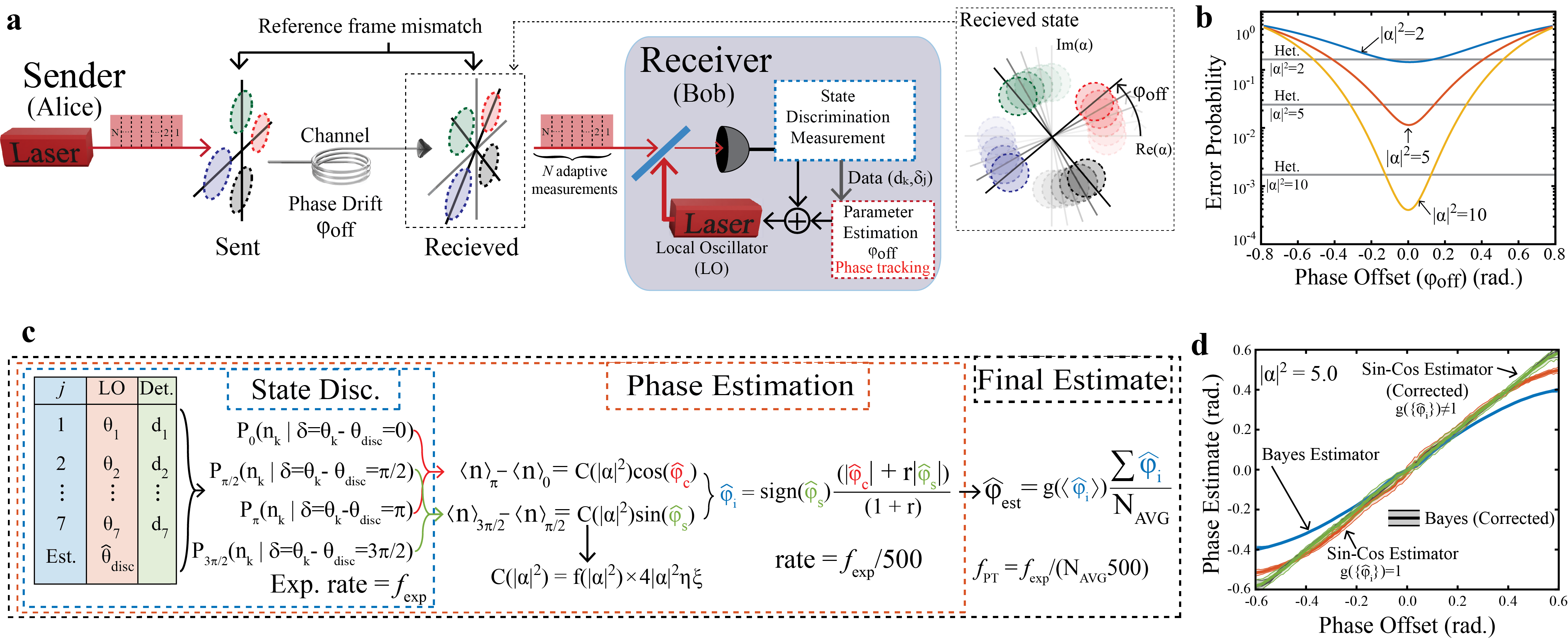}
	\caption{\label{A} \textbf{Phase tracking for non-Gaussian receivers surpassing the QNL} (a) A sender (Alice) prepares a coherent state $|\alpha_{k}\rangle$ with a phase $\theta_{k}\in\{0, \pi/2, \pi, 3\pi/2\}$ by phase modulation of lasers. The pulses propagate though a channel inducing random phase drifts $\phi_{\mathrm{off}}$. The receiver (Bob), uses a local oscillator (LO) to perform optimized discrimination non-Gaussian measurements \cite{becerra15}. The mismatch between the Alice's and Bob's phase reference frames caused by the channel increases the discrimination error for decoding information. (b) Probability of error for the adaptive non-Gaussian state discrimination measurement from Ref. \cite{becerra15} as a function of phase offset $\phi_{\mathrm{off}}$ between the signal and LO, for signal mean photon number $|\alpha|^{2}$ = 2.0, 5.0, and 10.0, together with the ideal heterodyne limit (Het.) for each mean photon number. (c) Flowchart of the algorithm followed for phase tracking. The discrimination measurement provides samples for phase estimation containing detection results $d_{j}$ for the relative phase between input state and the LO $\delta_{j}\in\{0,\pi/2,\pi,3\pi/2\}$. After 500 channel transmissions, the pairs $\{d_{j},\delta_{j}\}$ are used generate two estimates: $\hat{\phi}_{c}$ and $\hat{\phi}_{s}$. A weighted average $\hat{\phi}_{i}$ combines these estimates to increase accuracy. The final estimate $\hat{\phi}_{\mathrm{est}}$ is the average over $N_{\mathrm{avg}}$ phase estimates $\{\hat{\phi}_{i}\}$ multiplied by a gain function $g(\langle \hat{\phi}_{i}\rangle)$. The phase-tracking method feeds forward the estimate $\hat{\phi}_{\mathrm{est}}$ to the LO at a rate of $f_{\mathrm{PT}} = f_{\mathrm{expt}}/(500 N_{\mathrm{avg}})$ Hz for real-time phase tracking, with experimental repetition rate $f_{\mathrm{expt}}\approx12$ kHz. (d) Expected phase estimate as a function of applied phase for $|\alpha|^{2} = 5.0$ for the Sin-Cos estimator $\hat{\phi}_{\mathrm{est}}$ without (red line) and with (green line) optimized gain function $g(\langle\hat{\phi}_{i}\rangle)$, and for a Bayesian estimator (blue line), and the corrected Bayesian estimator (black line). The solid lines represent the mean of 100 Monte Carlo samples and shaded regions correspond to one standard deviation.
	}
	\label{concept}
\end{figure*}

Here we demonstrate a phase tracking method for non-Gaussian receivers for quadrature phase-shift-keyed (QPSK) coherent states \cite{becerra15} based on coherent displacement, adaptive measurements, and photon counting. The phase tracking method performs phase estimation and correction in real time using the data collected from the non-Gaussian discrimination measurement \cite{becerra15} without relying on strong phase-reference pilot pulses. This method enables the non-Gaussian receiver to overcome random phase variations encountered in realistic communication channels, while allowing the receiver to perform decoding measurements with sensitivities beyond the QNL, the shot-noise limit of conventional coherent receivers. This demonstration makes non-Gaussian receivers a more robust, feasible, and practical quantum technology for optical communications, and represents a significant advance for realizing low-power communications approaching the quantum limits in realistic communication channels.

\section{Phase tracking for Non-Gaussian Receivers} \label{PhasetrackingMethod}

Fig. \ref{concept}(a) shows the concept of phase tracking for a non-Gaussian measurement surpassing the QNL over a channel inducing random phase variations. The sender (Alice) uses laser pules to encode information in four coherent states with phases $\theta_{k} \in \{0, \pi/2, \pi, 3\pi/2\}$. The pulses propagate through the channel, which induces random phase shifts. The receiver (Bob) uses a laser as a local oscillator (LO) phase reference and performs a non-Gaussian discrimination measurement that surpasses the QNL for decoding the information \cite{becerra15}. The finite linewidths of the lasers in the transmitter and the receiver and the random channel phase variations cause mismatch between the phase space reference frames of Alice and Bob. These random phase drifts severely affect the expected performance of the state discrimination measurement. Figure \ref{concept}(b) shows the probability of error for the adaptive non-Gaussian measurement for discriminating four non-orthogonal states $|\alpha_{k}\rangle \in \{|\alpha\rangle,|i\alpha\rangle,|-\alpha\rangle,|-i\alpha\rangle\}$ below the heterodyne limit \cite{becerra15}, the QNL (see Appendix A), as a function of phase offset $\phi_{\mathrm{off}}$ between the input state $|\alpha_{k}\rangle$ and the receiver's LO, for mean photon numbers $|\alpha|^{2}$ = 2.0, 5.0, and 10.0. While the discrimination strategy demonstrated in \cite{becerra15} can tolerate small phase errors $\phi_{\mathrm{off}}$ without significant degradation, moderate values of $\phi_{\mathrm{off}}$ severely limit its performance, preventing discrimination below the QNL. To keep the expected performance benefit of the non-Gaussian measurement over the QNL, the receiver needs to perform phase tracking to correct for phase drifts induced by the channel.
While phase tracking based on heterodyne measurements can be realized with digital signal processing post measurement \cite{wang19, qi15, soh15}, non-Gaussian receivers require active phase tracking and correction in real time to maintain performances below the QNL \cite{becerra13, becerra15, ferdinand17}. Here we demonstrate a method for actively tracking and correcting for time-varying random phases for non-Gaussian receivers to enable sensitivities beyond the ideal heterodyne limit in channels inducing random phase variations.

The phase-tracking method for non-Gaussian receivers builds on a discrimination strategy to discriminate a state ${|\alpha_{k}\rangle\in\{|\alpha\rangle,|i\alpha\rangle,|-\alpha\rangle,|-i\alpha\rangle\}}$ implementing $N$ adaptive measurements with photon-number resolution \cite{becerra15}. During each adaptive measurement $j=1,2,...,N$, the receiver's LO performs hypothesis testing of the input state $|\alpha_{k}\rangle$ by adjusting its phase $\theta_{j}\in \{0, \pi/2, \pi, 3\pi/2\}$ according to a Bayesian discrimination strategy \cite{becerra15}. After $N$ adaptive measurements, the receiver provides an answer to the state discrimination problem $\theta_{\mathrm{disc}}$ about the phase of the input state $|\alpha_{k}\rangle$ (see Appendix {\color{blue}\ref{AppStDiscStr}}). Assuming that this answer $\theta_{\mathrm{disc}}$  is correct, the data collected during the $N$ adaptive measurements can now be used as samples for estimation of the phase $\phi_{\mathrm{off}}$ induced by the channel. For a discrimination measurement, these data consists of $N$ photon-counting detections $\{d_{1},d_{2},...,d_{N}\}$, together with the LO's phases during each adaptive measurement $\{\theta_{1},\theta_{2},...,\theta_{N}\}$. Since the answer to the discrimination problem $\theta_{\mathrm{disc}}$ is a very good estimate of the phase of the input state, it can be used to estimate the relative phases $\delta_{j}$ between the input state and the LO in each adaptive measurement as $\delta_{j}=\theta_{j}-\theta_{\mathrm{disc}}$. The data for estimating $\phi_{\mathrm{off}}$ then consist of the pairs $\{d_{j}, \delta_{j}\}$. Accumulating data during a moderate number of channel transmissions allows for estimating $\phi_{\mathrm{off}}$ in real time and performing phase tracking simultaneously with the state discrimination measurement \cite{becerra15}. This method enables the receiver to utilize the data from the discrimination measurement to estimate and correct for random phase excursions.

To obtain an estimate of $\phi_{\mathrm{off}}$, the phase tracking method uses the collected data $\{d_{j}, \delta_{j}\}$ from discrimination measurements over 500 channel transmissions. This data consists of photon counting samples of the interference between the input state and the LO for relative phases $\delta = \{0, \pi/2, \pi, 3\pi/2\}$. In principle, there are different estimators that can produce an estimate from the pairs $\{d_{j}, \delta_{j}\}$ (see Appendix {\color{blue}\ref{AppPhEstPerf}} for two possible estimators). However, phase tracking for non-Gaussian receivers requires a simple estimator that can be efficiently calculated in real time, while being robust to the unavoidable errors from the discrimination measurement. A simple estimator can be obtained by using $\{d_{j}, \delta_{j}\}$ to generate four photon number (Poisson) distributions $P_{0}(n_{k}|\delta=0)$, $P_{\pi/2}(n_{k}|\delta=\pi/2)$, $P_{\pi}(n_{k}|\delta=\pi)$, and $P_{3\pi/2}(n_{k}|\delta=3\pi/2)$ for $\delta = \{0, \pi/2, \pi, 3\pi/2\}$ (see Fig \ref{concept}(c)). Here $n_{k}$ is the photon number of detected photons for different distributions. By calculating the differences between means $\langle n \rangle_{\delta}$ of these distributions we can form estimates of $\phi_{\mathrm{off}}$ as:
\begin{align}
 \langle n \rangle_{\pi} - \langle n \rangle_{0} &= C(|\alpha|^{2}) \textrm{cos}(\hat{\phi}_{c})
 \\
 \nonumber
 \\
  \langle n \rangle_{3\pi/2} - \langle n \rangle_{\pi/2} &= C(|\alpha|^{2}) \textrm{sin}(\hat{\phi}_{s})
 \end{align}
with
 \begin{align}
 C(|\alpha|^{2}) &= f(|\alpha|^{2}) \times 4|\alpha|^{2}\eta \xi/N
 \end{align}
where $\eta$ is the detection efficiency, $\xi$ is the interference visibility, and $\hat{\phi}_{c,s}$ are the phase estimates. $f(|\alpha|^{2})$ is a factor that is used to reduce a bias in the phase estimates arising from the non-zero probability of error for the state discrimination strategy (see Appendix {\color{blue}\ref{AppPhEstPerf}}).
Errors in the state discrimination measurement ($P_{\mathrm{E}}\neq0$) cause errors in populating the distributions $P_{\delta}$ and thus in their mean values $\langle n \rangle_{\delta}$. These errors cause $\hat{\phi}_{c}$ to be biased away from zero by making $\langle n \rangle_{\pi} - \langle n \rangle_{0} < 4|\alpha|^{2}\eta \xi/N$. The function $f(|\alpha|^{2})$ allows for correcting these biases by making $\langle n \rangle_{\pi} - \langle n \rangle_{0} \approx f(|\alpha|^{2}) \times 4|\alpha|^{2}\eta \xi/N$, thus reducing the effects of discrimination errors in the phase estimation procedure. These errors also produce biases in $\hat{\phi}_{s}$ by causing $\langle n \rangle_{3\pi/2} - \langle n \rangle_{\pi/2}$ to be reduced. Appendix \ref{AppPhEstPerf} 1 describes the procedure to obtain the optimal values of $f(|\alpha|^{2})$ for reducing the effects of discrimination errors. As a second step, the phase-tracking method uses a weighted average of estimates $\hat{\phi}_{c}$ and $\hat{\phi}_{s}$ with relative weight $r$  to obtain an estimate $\hat{\phi}_{i}$ of $\phi_{\mathrm{off}}$ within 500 transmissions (see Fig. \ref{concept}(b)). The weight $r$, determined from Monte Carlo simulations, allows for reducing the difference between $\hat{\phi}_{i}$ and $\phi_{\mathrm{off}}$  at the end points of the capture range for phase tracking ($\pm0.6$ rad in our experimental demonstration). The final estimate $\hat{\phi}_{\mathrm{est}}$ in the phase tracking method is obtained by averaging over $N_{\textrm{avg}}$ estimates $\hat{\phi}_{i}$ and multiplying by a gain factor $g(\langle\hat{\phi}_{i}\rangle)$  that depends on the $N_{\textrm{avg}}$  estimates $\{\hat{\phi}_{i}\}$ (see Appendix {\color{blue}\ref{AppEstPerf-Nave}}). This gain factor reduces the difference between the applied phase $\phi_{\mathrm{off}}$ and the estimated phases $\{\hat{\phi}_{i}\}$. Figure \ref{concept} (d) shows the result of Monte Carlo simulations  of the final phase estimate $\hat{\phi}_{\mathrm{est}}$ with $g(\langle\hat{\phi}_{i}\rangle)=1$ (red line) and $g(\langle\hat{\phi}_{i}\rangle)$ optimized to approach the true phase $\phi_{\mathrm{off}}$ (green line). This final estimate $\hat{\phi}_{\mathrm{est}}$  is used to feed forward to the receiver's LO every 500$\times N_{\mathrm{avg}}$ channel transmissions at a rate $f_{\mathrm{PT}}$ for phase drift correction. This method enables real-time phase tracking for correcting time varying phases while enabling the non-Gaussian receiver to surpass the QNL.

\begin{figure}[!tp]
\centering\includegraphics[width = 8.5cm]{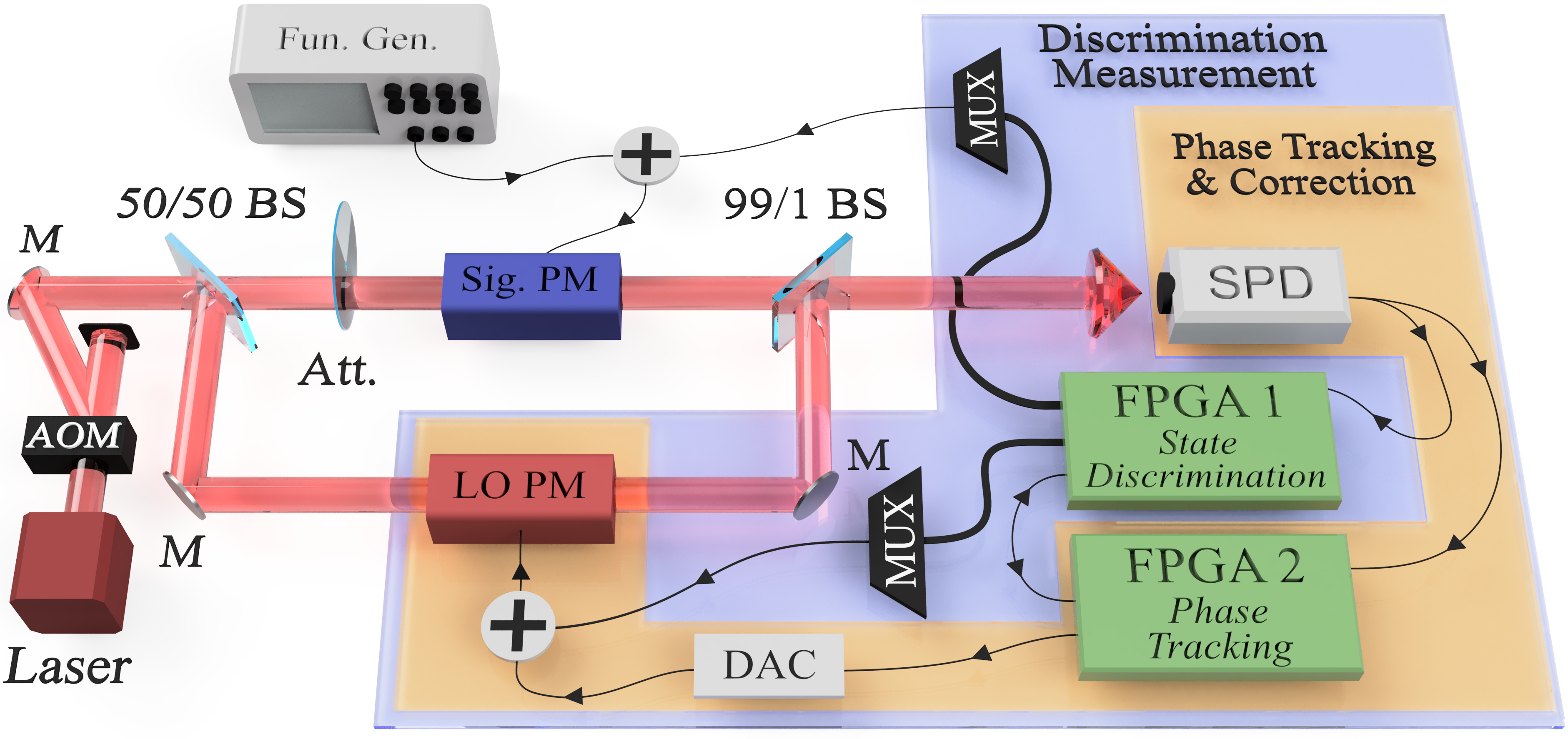}
\caption{\label{A} \textbf{Experimental Configuration} Experimental setup for the demonstration of phase tracking for non-Gaussian state discrimination measurements for QPSK states surpassing the QNL. See main text for details. AOM, Acousto-optic modulator; Att., variable attenuator; PM, phase modulator; MUX, multiplexer; FPGA, field-programmable gate array; SPD, single photon detector; Fun. Gen, function generator; M, Mirror.}
\label{ExpConfig}
\end{figure}

\section{Experimental configuration}

Figure \ref{ExpConfig} shows the experimental configuration for the demonstration of phase tracking of adaptive non-Gaussian state discrimination measurements for QPSK states $\{|\alpha\rangle,|i\alpha\rangle,|-\alpha\rangle,|-i\alpha\rangle\}$. The measurement strategy consists of $N=7$ adaptive measurements via feedback in the phase of the LO \cite{becerra15}. A helium-neon (HeNe) laser at 633 nm and an acousto-optic modulator (AOM) prepare 35-$\mu s$ coherent state pulses at a rate of $f_{\mathrm{expt}}\approx12$ kHz. The light pulses enter an unbalanced Mach-Zender interferometer through a 50/50 beam splitter. We prepare the phases of the input signal state and the LO with two 4:1 multiplexers (MUX) and two phase modulators (PM). The input states and LO interfere in a 99/1 beam splitter, which implements the displacement operation of the input state \cite{paris96}. A field programmable gate array (FPGA), FPGA1 in Fig. \ref{ExpConfig}, implements the discrimination strategy based on adaptive measurements and photon number resolving (PNR) detection described in Ref. \cite{becerra15}. This FPGA1 controls the timing of the experiment, processes photon detections, and updates the phase of the LO for each adaptive measurement. The overall detection efficiency of the experiment is $\eta=72\%$, with interference visibility $\xi=99.8\%$.

We use a second FPGA (FPGA2) to perform active phase tracking using the data pairs $\{d_{j}, \delta_{j}\}$ sent from FPGA1 generated from the state discrimination measurement, as described above. FPGA2 performs real-time phase estimation to obtain an estimate of $\phi_{\mathrm{off}}$, and feeds forward this information to the receiver to adjust the LO phase to perform phase tracking and correction. Controllable phase offsets and phase noise in the input state are prepared with an arbitrary waveform function generator (FG) \cite{dimario19} for investigating the phase tracking method in channels inducing random phase variations. We use an 8-bit Digital to Analog Converter (DAC) to feed forward the estimated phase offset to the LO. We chose a finite capture range $R$ of phases for feed forward to the phase of the LO equal to $R=[-0.6,+0.6]$ rad. This choice results in a phase resolution of about 1.2 rad$ / 2^{8} = 5$ mrad for phase tracking, while having a large enough capture range.

The absolute power of the input state is calibrated using a photodiode-based light-trapping (TRAP) detector with a 0.05\% uncertainty tied to an absolute spectral response scale \cite{gentile96}. This TRAP detector was used to calibrate a series of attenuators to lower the power of a power-stabilized 633 nm laser to the single-photon level with a combined 1$\sigma$ uncertainty of 1.8\%, and the transmission of the optical elements from where the state is prepared to where it is detected with transmittance $T$= 92.5(2)$\%$. This results in a total uncertainty for the calibration of the absolute average photon number per pulse of $\sigma \approx 2 \%$. The FPGAs used for implementing the state discrimination strategy and phase tracking were both Altera Cyclone II FPGAs, model EP2C5T144C8 with 4608 logical elements, base clock of 48 MHz, and 158 digital I/O pins.

\section{Results}
We investigate the performance of the phase-tracking method under different scenarios. In the first scenario, the input state experiences a sudden constant phase offset and the phase tracking method needs to estimate and correct for large phase offsets. The second scenario aims to simulate a realistic channel inducing time-varying phase noise, where the input state experiences Gaussian random walks in phase at different diffusion rates. This scenario allows us to investigate phase-tracking of random phase drifts in the channel and the impact of tracking bandwidth on the performance of non-Gaussian receivers.

\subsection{Phase tracking under constant phase offsets}
\begin{figure}[!tbp]
\centering\includegraphics[width = 7cm]{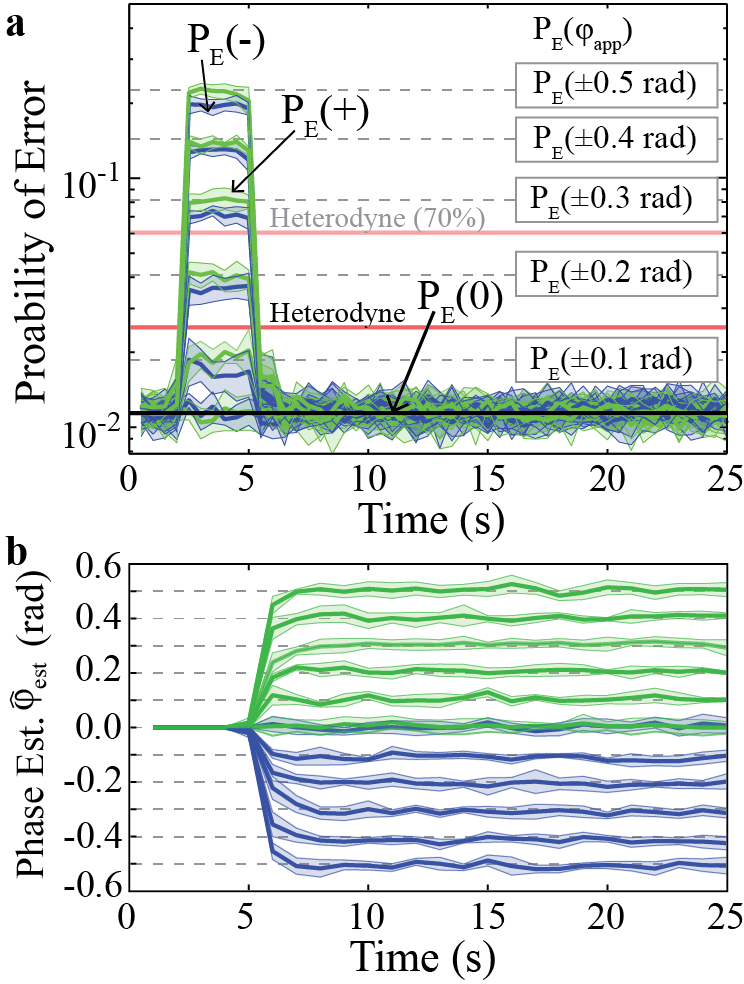}
\caption{\label{A} \textbf{Phase tracking under constant phase offsets} (a) Probability of error $P_{\mathrm{E}}$ as a function of time ($t$) shown for every $0.5$s time bin. From $t$=0 to 2s no phase offset is applied, $\phi_{\mathrm{app}}=0$. At $t$=2s, the input state experiences  a constant phase offset $\phi_{\mathrm{app}}$ producing an increase in $P_{\textrm{E}}$. At $t$=4 s, the phase tracking turns on and corrects for $\phi_{\mathrm{app}}$. (b) Phase tracking estimates $\hat{\phi}_{\mathrm{est}}$ as a function of time estimated every $T_{\mathrm{PT}}\approx 0.85$ s. Dashed lines show expected error probabilities $P_{\textrm{E}}$ in (a) and applied phase offsets $\phi_{\mathrm{app}}$ in (b).}
\label{const}
\end{figure}

Figure \ref{const} shows the performance of the phase tracking method under sudden constant phase offsets of $\phi_{\mathrm{app}}$=$\{\pm0.1, \pm0.2, \pm0.3, \pm0.4, \pm0.5\}$ rad of the input state with mean photon number $|\alpha|^{2} = 5.0$. Figure \ref{const}(a) shows the probability of error $P_{\textrm{E}}$ calculated for time bins of about $0.5$ s. Figure \ref{const}(b) shows the phase estimates $\hat{\phi}_{\mathrm{est}}$ generated by the phase-tracking method as a function of time. Thick lines represent the average over 5 independent experimental runs, each time bin corresponds to about $5\times10^3$ ($\approx f_{\mathrm{exp}}\times0.5$ s) independent experiments, and shaded regions represent one standard deviation. Green (blue) lines correspond to positive (negative) applied phase offsets $\phi_{\mathrm{app}}$. The phase-tracking method here uses $N_{\textrm{avg}} = 20$, so that each estimate $\hat{\phi}_{\mathrm{est}}$ was obtained at a phase tracking rate $f_{\mathrm{PT}}=1/T_{\mathrm{PT}}\approx 1/ 0.85$ s [See Fig. \ref{concept}(c)]. For this investigation, the relative phase of the signal and LO was locked between each experimental trial, similar to Ref. \cite{becerra15}, and the phase offset $\phi_{\mathrm{app}}$ was applied during each state discrimination measurement.
\begin{figure*}[!tbp]
\centering\includegraphics[width =\textwidth]{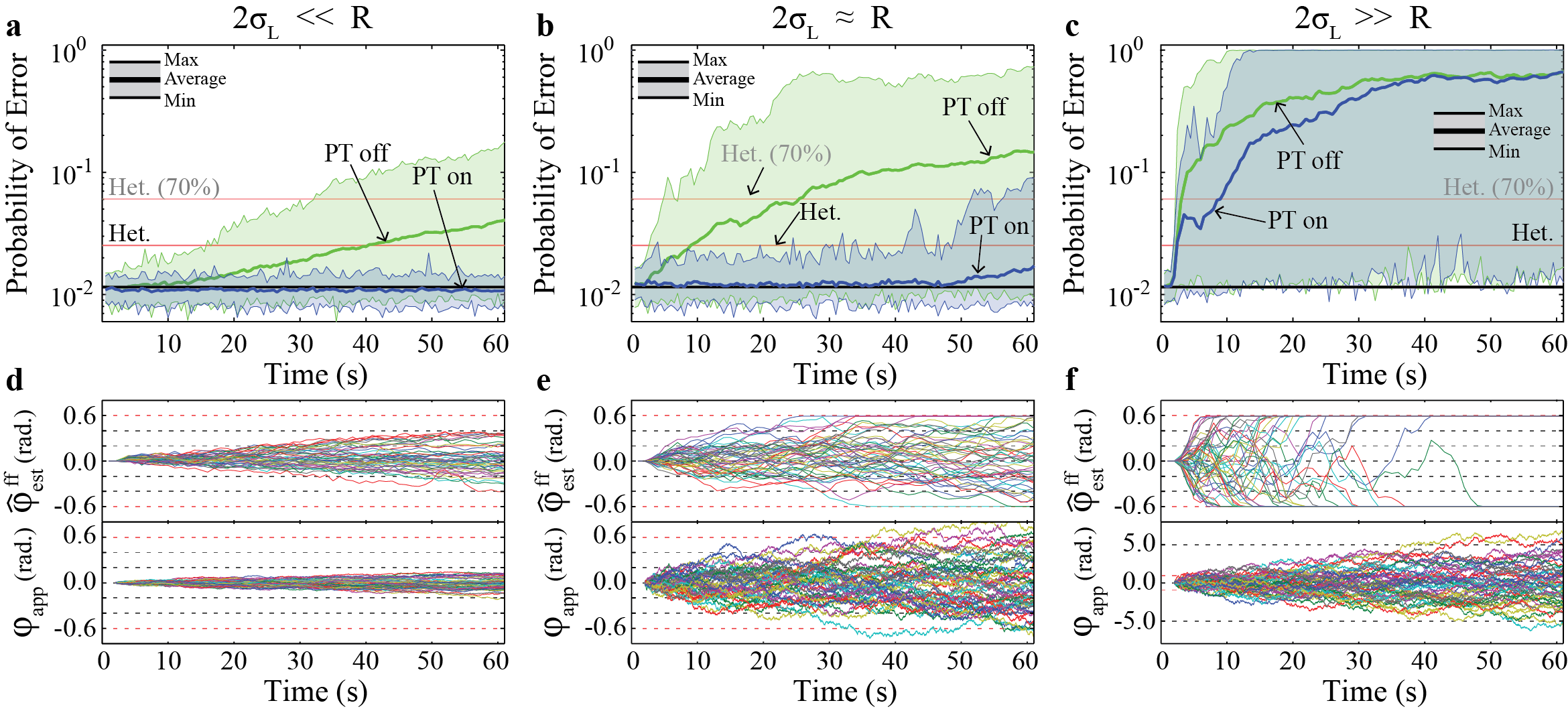}
\caption{\label{A} \textbf{Phase-tracking (PT) of Gaussian random walks in phase.}
(a)--(c) Probability of error $P_{\textrm{E}}$  for $|\alpha|^{2} = 5.0$ as a function of time for 50 realizations of Gaussian random walks  in phase applied to the input signal for: (a) $\sigma_{1}=1$ mrad $(=0.2\times R/2\sqrt{L})$, (b) $\sigma_{1}=5$ mrad $(=1\times R/2\sqrt{L})$, and (c) $\sigma_{1}=5$ mrad $(=25\times R/2\sqrt{L})$, with total variances $\sigma_{L}^2=L\sigma_{1}^2$ for each case. Here, $L=6500$ corresponds to the total steps in the random walks, and $R=[-0.6,0.6]$ rad is the capture range of phase tracking in our experiment. Bold lines show the average and shade regions the spread for these walks with (blue) and without (green) phase tracking. (d)--(f) Applied phases $\phi_{\mathrm{app}}$ during Gaussian random walks and the phase estimates $\hat{\phi}_{\mathrm{est}}^{ff}=\hat{\phi}_{\mathrm{est}}$ generated by the phase tracking method for cases (a)--(c), respectively. The phase estimates $\hat{\phi}_{\mathrm{est}}^{ff}$ used to feed forward to the LO for phase correction are bounded by the experimental capture range $R$, as can be seen in (e) and (f).}
\label{mpn5}
\end{figure*}

From $t=0$ to 2s, we verify that the performance of the receiver is 3.4 dB below the heterodyne limit (red line) without the applied phase offset $\phi_{\mathrm{app}}=0$. At $t=2$s, we apply a constant phase offset $\phi_{\mathrm{app}}$ to the input state, producing sudden jumps in the probability of error $P_{\textrm{E}}$ depending on the value of $\phi_{\mathrm{app}}$. At $t=4$ s, the phase tracking method is turned on. After an estimation cycle $T_{\mathrm{PT}}\approx 0.85$ s, the phase tracking produces an estimate $\hat{\phi}_{\mathrm{est}}$ and corrects for $\phi_{\mathrm{app}}$, allowing the receiver to perform below the heterodyne limit for all phase offsets. Figure \ref{const}(b) shows the phase estimates $\hat{\phi}_{\mathrm{est}}$ as a function of time, demonstrating that the phase-tracking method accurately identifies and corrects for large phase offsets. We observe that this method enables the non-Gaussian receiver to keep its expected advantage of 3.4 dB over an ideal heterodyne measurement.

\subsection{Phase tracking of random walks in phase}

\subsubsection{Phase tracking with different noise strengths}

In coherent optical communications, the receiver is usually required to decode information encoded in coherent state signals in the presence of time-dependent random variations in phase, which severely limits the receiver's ability to recover the information. We investigate the phase-tracking method for situations where the input state experiences Gaussian phase noise \cite{ip07}, which could include effects of phase noise in the LO and the transmitter laser or random fluctuations arising from different processes \cite{kikuchi08,ip07,xie11,lygagnon06,ghozlan13,khanzadi15}.
Gaussian random walks in phase have been broadly used as an acceptable model for phase noise in optical communications and phase drift between the sender and receiver in classical \cite{lygagnon06,goldfarb06,salz86, kikuchi16} and quantum communications \cite{wang19, qi15,wang19}. We note that the algorithm used for phase tracking is independent of the choice of phase noise model, as it makes no assumptions about the dynamics of the noise and the noise model is not used to obtain the phase estimate (see Sec. \ref{PhasetrackingMethod}).

For this study, we do not stabilize the relative phase between the input signal state and the LO. Under these conditions, the receiver experiences the drift of the experimental setup and induced random walks in phase. This situation is analogous to having a LO whose phase is constantly drifting  and a channel that induces phase noise. This experimental configuration aims to mimic more realistic situations where the signal and LO are generated from different lasers \cite{ip07}. In this investigation, the phase noise in the signal is implemented by preparing discrete Gaussian random walks in phase with $L$=6500 steps, each distributed according to a zero-mean Gaussian distribution with standard deviation $\sigma_{1}$ \cite{ip07}.

Figures \ref{mpn5}(a)-(c) show the probability of error $P_{\textrm{E}}$ for $|\alpha|^{2} = 5.0$ as a function of time for 50 realizations of discrete Gaussian random walks in phase at a rate of $f_{\mathrm{RW}} = 100$ Hz for walks with (a) $\sigma_{1}=0.1$ mrad, (b) $\sigma_{1}=5$ mrad, and (c) $\sigma_{1}=25$ mrad. Thick lines show the average over 50 walks and shaded regions represent the spread for these walks with (blue) and without (green) phase tracking (PT) with $N_{\textrm{avg}} = 20$, so that $f_{\mathrm{PT}}\approx 1.2$ Hz. After $L$ steps the total variance of the Gaussian random walks are $\sigma_{L}^{2}=L\sigma_{1}^{2}$, so that situations in Figs. \ref{mpn5}(a)--\ref{mpn5}(c) correspond to different regimes: (a) small ($2\sigma_{L}< R$), (b) moderate ($2\sigma_{L}\approx R$), and (c) severe ($2\sigma_{L}\gg R$) phase noise. Here, $R=[-0.6,0.6]$ rad is the capture range for phase tracking in our experiment, which is experimentally chosen. Figures \ref{mpn5}(d)--\ref{mpn5}(f) show the applied phase $\phi_{\mathrm{app}}$ during the Gaussian random walks for cases in Figs. \ref{mpn5}(a)--\ref{mpn5}(c), respectively, and the phase estimate $\hat{\phi}_{\mathrm{est}}^{ff}$ that is the actual phase sent to the LO for phase tracking via feed forward. Note that the forwarded phase $\hat{\phi}_{\mathrm{est}}^{ff}$ is restricted to be within the experimental capture range $R$ so that $\hat{\phi}_{\mathrm{est}}^{ff}=\hat{\phi}_{\mathrm{est}}$ under the condition $|\hat{\phi}_{\mathrm{est}}^{ff}|<|R|$.

We observe that, in general, Gaussian random walks in phase severely degrade the performance of the non-Gaussian receiver precluding any advantage over the heterodyne limit. However, in situations with small ($2\sigma_{L}< R$) and moderate ($2\sigma_{L}\approx R$) levels of noise in Figs. \ref{mpn5}(a) and  (b), respectively, the phase-tracking method accurately estimates and corrects for phase noise, enabling the receiver to maintain its performance 3.4 dB below the heterodyne limit. For small phase noise in Fig. \ref{mpn5}(d) with $2\sigma_{L}< R$, the applied phase $\phi_{\mathrm{app}}$ is smaller than the phase drifts in the experiment. The estimated phase $\hat{\phi}_{\mathrm{est}}^{ff}$ captures the contributions of $\phi_{\mathrm{app}}$ and of these drifts showing a larger variance than $\phi_{\mathrm{app}}$. For situations with moderate phase noise with $2\sigma_{L} \approx R$, the applied walks in phase $\phi_{\mathrm{app}}$ contain walks that exceed the capture range $R$ at some point in time, as shown in Fig. \ref{mpn5}(e). After this point, the estimated phases $\hat{\phi}_{\mathrm{est}}$ for these walks are clamped at $|R|$ to generate $\hat{\phi}_{\mathrm{est}}^{ff}$ to feed forward to the LO. This procedure produces a slight increase in $P_{\textrm{E}}$ after 50s, as shown in Fig. \ref{mpn5}(b).

For situations with large phase noise in Fig. \ref{mpn5}(c) for which $2\sigma_{L}\gg R$, the receiver's performance degrades above the heterodyne limit within a short time. The phase tracking method can reduce the effects of phase noise. However, since the applied phases $\phi_{\mathrm{app}}$ rapidly exceed the capture range $R$, the estimated phases $\hat{\phi}_{\mathrm{est}}^{ff}$  that are fed forward to the LO are clamped at $R$ for many cases, as can be seen in Fig. \ref{mpn5}(f). This procedure limits the performance for phase tracking in our current implementation. However, increasing the resolution of the electronic controller {\color{black}and the DAC to 10-bits} used to feed forward to the LO phase can allow for increasing the capture range to $R=\pm\pi$ rad, while maintaining good phase resolution {\color{black}of $\approx6$ mrad} for phase tracking. In this case, whenever $\phi_{\mathrm{app}}$ reaches this range, the estimate $\hat{\phi}_{\mathrm{est}}^{ff}$ would wrap around from $\pm\pi$ to $\mp\pi$. This procedure would allow for tracking phase walks exceeding $R$ and maintaining the receiver's performance below the heterodyne limit under any level of phase noise.

\subsubsection{Phase tracking with different input powers}

The performance of the phase-tracking method for non-Gaussian receivers critically depends on the performance of the state discrimination measurement. The information used for phase estimation and tracking $\{d_{j}, \delta_{j}\}$ assumes the answer to the discrimination problem ${\theta}_{\mathrm{disc}}$ to be correct, which is true only with probability $P_{\textrm{C}} = 1-P_{\textrm{E}}$. Since $P_{\textrm{E}}$ depends strongly on the mean photon number $|\alpha|^{2}$ of the input state \cite{becerra15}, the receiver's ability to perform phase tracking will also depend on $|\alpha|^{2}$. Larger input powers $|\alpha|^{2}$ result in lower error probabilities $P_{\textrm{E}}$, and can allow the phase-tracking method to perform phase estimation with higher accuracy, achieve higher tracking rates $f_{\mathrm{PT}}$, and correct for phase noise with higher bandwidths $f_{\mathrm{RW}}$. On the other hand, for low powers $|\alpha|^{2}$ the performance of the phase tracking method is affected due to higher $P_{\textrm{E}}$. However, achieving phase tracking in these two power regimes is required for both low-power classical \cite{lygagnon06, ip07, kikuchi08, ip08} and quantum \cite{grosshans03, qi15, soh15} communications.

\begin{figure}[!bp]
\centering\includegraphics[width = 8.5cm]{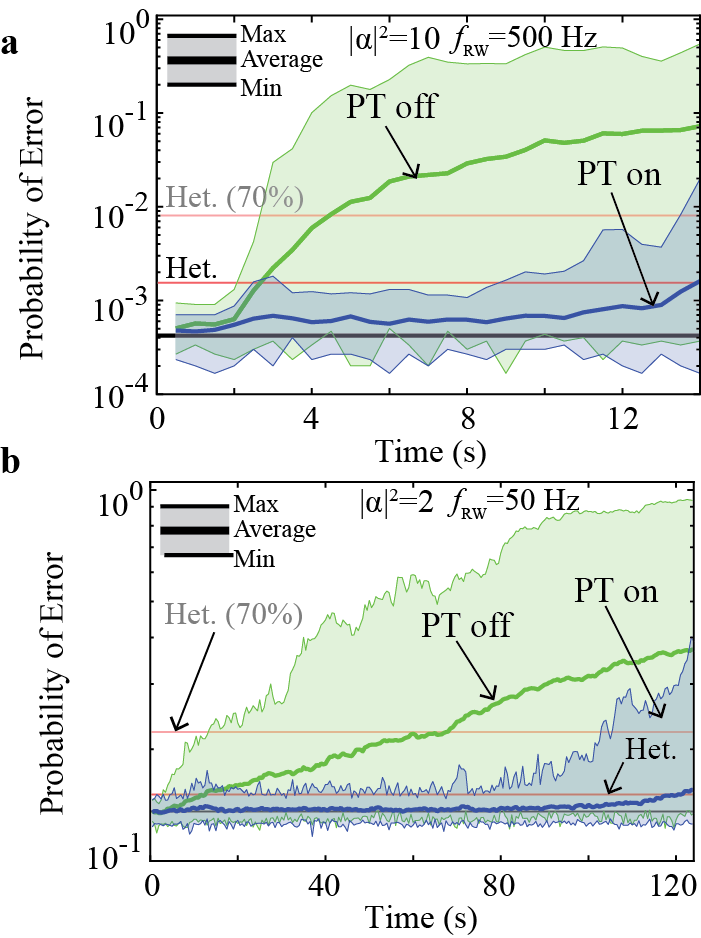}
\caption{\label{A} \textbf{Phase tracking with different mean photon numbers} Probability of error as a function of time under Gaussian phase noise with (blue) and without (green) phase tracking for (a) $|\alpha|^{2} = 10.0$ and (b) $|\alpha|^{2} = 2.0$. Thick lines show the averages and shaded regions show the spread in $P_{\textrm{E}}$ over 50 Gaussian random walks. Parameters for phase tracking, $f_{\mathrm{PT}}$, and for the Gaussian random walks, $f_{\mathrm{RW}}$ and $\sigma_{i}$, are chosen to satisfy $2\sigma_{L}\approx R$, analogous to the situation in Fig. \ref{mpn5} (b) for $|\alpha|^{2} = 5.0$. Black lines show the expected performance of the receiver in the absence of Gaussian phase noise: 6 dB and 0.45 dB below the heterodyne limit for $|\alpha|^{2} = 10.0$ and $|\alpha|^{2} = 2.0$, respectively \cite{becerra15}. Note that the phase tracking method enables the receiver to perform below the heterodyne limit under Gaussian phase noise in the high- and low-input power regimes.}
\label{mpn10}
\end{figure}

Figure \ref{mpn10} shows the performance of the phase-tracking method for $|\alpha|^{2} = 10.0$ (a) and (b) $|\alpha|^{2} = 2.0$. phase-tracking in these high- and low-input power regimes can be implemented at different rates $f_{\mathrm{PT}}$  to correct for noise with different strengths and bandwidths. In the two plots in Fig. \ref{mpn10}, the phase-tracking parameters and Gaussian phase noise are chosen to satisfy the condition $2\sigma_{L} =2\sqrt{L}\sigma_{1} = 2f_{\mathrm{RW}}T\sigma_{1} \approx R$, so that these situations are analogous to the one shown in Fig. \ref{mpn5} (b) for $|\alpha|^{2} = 5.0$. Here $\sigma_{1}=5$ mrad, and $T$ is the displayed period: $T=13$ s for $|\alpha|^{2}=10.0$; and $T=120$ s for $|\alpha|^{2} = 2.0$. For $|\alpha|^{2} = 10.0$, phase-tracking achieves higher estimation accuracy, and can be reliably implemented with $N_{\textrm{avg}} = 4$, enabling phase-tracking rates of $f_{\mathrm{PT}} \approx 5.8$ Hz, five times faster than for $|\alpha|^{2} = 5.0$. As a result, the receiver can track and correct for random Gaussian phase noise with a rate of $f_{\mathrm{RW}} = 500$ Hz, while performing below the heterodyne limit (See Fig. \ref{mpn10}(a)). phase-tracking for $|\alpha|^{2} = 2.0$ requires more samples to obtain accurate phase estimates, and can be implemented reliably with $N_{\textrm{avg}} = 40$ with a tracking rate $f_{\mathrm{PT}} \approx 0.5$ Hz. In this case, the receiver can track and correct for phase noise at a rate of $f_{\mathrm{RW}} = 50$ Hz, while performing below the heterodyne limit [See Fig. \ref{mpn10}(b)].

\subsubsection{phase-tracking of noise with different bandwidths}
The phase-tracking method has a strong dependence on the noise bandwidth present in the communication channel and how it compares to the rate at which phase-tracking can be implemented \cite{ip07}. We have studied the performance of the phase-tracking method for non-Gaussian receivers for tracking random phase noise with different noise bandwidths. This study is described in Appendix \ref{AppEstPerf-BW}. In our findings we observe that for a fixed $f_{\mathrm{PT}}$, the phase-tracking method can correct for noise with different bandwidths $f_{\mathrm{RW}}$. We note, however, that $f_{\mathrm{PT}}$ has to be high enough to keep the receiver's performance below the heterodyne limit for extended periods of time. As one example, we observe that for a non-Gaussian receiver with $|\alpha|^{2} = 5.0$ and $f_{\mathrm{PT}} = 1.2$ Hz, reliable phase-tracking of random noise can be performed for noise bandwidths $f_{\mathrm{RW}} = 100$ Hz, and sub-QNL sensitivity can be kept for $f_{\mathrm{RW}} = 500$ Hz for $\approx$ 10 s. Tracking noise with higher bandwidths can be achieved with higher experimental rates $f_{\mathrm{expt}}>11$ kHz to increase $f_{\mathrm{PT}}$ or with larger mean photon numbers $|\alpha|^{2}$ to generate more accurate phase estimates  $\hat{\phi}_{\mathrm{est}}$.

\section{Discussion}

Receivers with sensitivities surpassing the QNL of ideal conventional receivers have a large potential for enabling efficient and reliable low-power communications at the single- and few-photon levels. The phase-tracking method demonstrated here for non-Gaussian receivers allows for tracking random phase variations and noise with different strengths and bandwidths. This method provides the much needed robustness to enable non-Gaussian receivers to perform below the QNL in channels with phase noise for a wide range of powers. We note that the phase-tracking bandwidth in our proof-of-principle demonstration was implemented at low rates because of experimental constraints, and used a single laser shared between transmitter and receiver. However, using an estimator with higher estimation accuracy, such as the Bayesian estimator, combined with high-bandwidth electronics and efficient single-photon detectors \cite{holzman2019}, would allow the receiver's measurement and phase-tracking to be realized at much higher bandwidths. This in turn would enable non-Gaussian receivers to overcome realistic noise in communication channels with independent lasers at the receiver and transmitter \cite{ip08}, while outperforming ideal shot-noise-limited coherent receivers \cite{kikuchi16}.

We note that transmissions of high-power pulses interleaved with the input states could be used with a heterodyne detection for phase-tracking \cite{qi07, soh15}, without relying on knowledge of the power of the input coherent states to be discriminated and the visibility of the interference with the LO. The phase-tracking method presented here uses only the data directly collected from the non-Gaussian measurement that assumes a known intensity and visibility. However, the data from the state discrimination strategy could in principle be used to estimate the input intensity and visibility in addition to the phase offset, and allow for tracking of multiple time-varying parameters without the need for dedicated light pulses for estimation and tracking. We also note that it may be possible to split the power of the input state to use a fraction of light to perform phase estimation with a heterodyne measurement. However, the estimation precision of these split-and-estimate methods for phase-tracking will depend on the fraction of power used for phase estimation, and there will be an increase in the probability of error in the state discrimination due to the reduced power entering the sub-shot-noise receiver.

In the future, enabling coherent communication technologies that can approach the quantum limits in sensitivity and information transfer in realistic channels at low powers will require the ability to track other impairments in the channel including polarization rotation, background noise, and power variations. While we demonstrated a method for tracking to correct phase drifts induced by a channel, we believe that the data from the state discrimination measurement that are used for phase-tracking can be leveraged for estimation and tracking of other sources of noise in the channel, such as amplitude noise (see Appendix E). Moreover, we anticipate that this technique for phase-tracking can be applied to optimized non-Gaussian measurements surpassing the QNL in the single-photon regime \cite{ferdinand17}. This possibility can enable phase-tracking in quantum key distribution for secure communications at very low powers without requiring strong phase reference pilot pulses \cite{qi15}.

\section{Conclusions}

We demonstrate a phase-tracking method for non-Gaussian receivers \cite{becerra15} for phase-encoded coherent states surpassing the sensitivity limits of shot-noise limited coherent receivers: the quantum noise limit (QNL) \cite{kikuchi16}. The phase-tracking method performs phase estimation and correction in real time using the data from the non-Gaussian discrimination measurement \cite{becerra15}, without continuously relying on phase reference pilot fields from the transmitter. Our experimental demonstration shows that the phase-tracking method provides non-Gaussian receivers with the required robustness to overcome random phase noise encountered in realistic communication channels, and enables the receiver to perform measurements beyond the QNL under diverse conditions with different noise strengths and bandwidths. Moreover, since the phase-tracking method uses the data from a measurement surpassing the QNL at very low power levels, this method is well suited for assisting quantum communication protocols based on weak coherent states for efficient \cite{xu15, guan16} and secure \cite{arrazola14, clarke12, bennett92, huttner95, grosshans03, takeoka14, pirandola17, ghorai19} communications. Our demonstration of phase-tracking for non-Gaussian receivers makes sub-shot-noise-limited receivers a more robust, feasible, and practical quantum technology for low-power communications based on coherent states for approaching the quantum limits in realistic communication channels.
\\
\\
\noindent
\textbf{ACKNOWLEDGEMENTS}
\\
This work was supported by the National Science Foundation (NSF) (PHY-1653670, PHY-1521016).
\\
\\

\appendix

\section{Quantum Noise Limit}
The Quantum Noise Limit (QNL) for the discrimination of coherent states in a given encoding scheme is obtained through the probability of error in discrimination:
\begin{equation}
\mathrm{P}_{\mathrm{E}} = 1 - \mathrm{P}_{\mathrm{C}} = 1 - \sum\limits_{k=1}^{M} \mathrm{P}(\alpha_{k}) \mathrm{P}(\alpha_{k} | \alpha_{k})
\label{error_prob}
\end{equation}
where $M$ is the number of states in the alphabet, and $\mathrm{P}(\alpha_{k})$ is the prior probability of state $|\alpha_{k}\rangle$ which is equal to $\frac{1}{M}$ for equiprobable states. $\mathrm{P}(\alpha_{k} | \alpha_{k})$ is the probability of guessing state $|\alpha_{k}\rangle$ given that state $|\alpha_{k}\rangle$ was sent, i.e. the probability of correct discrimination.

For the discrimination of two coherent states in the binary phase shift keying (BPSK) format $| \alpha_{k} \rangle \in \{| \pm \alpha \rangle \}$, the homodyne measurement along the $x$-quadrature is the optimal Gaussian measurement \cite{takeoka08}. The probability of error for the homodyne measurement corresponds to the QNL for the BPSK alphabet. $\mathrm{P}(\alpha_{k} | \alpha_{k})$ for a homodyne measurement is \cite{proakis00}:
\begin{align}
\mathrm{P}(\alpha_{k} | \alpha_{k}) &= \frac{1}{\sqrt{\pi}} \int\limits_{R} e^{-(x - \sqrt{2}\alpha_{k})^{2}}dx
\nonumber
\\
&= \frac{1}{2}\big(1 + \mathrm{erf}(\sqrt{2}\alpha)\big)
\end{align}
where $R$ is the region where $| \alpha_{k} \rangle$ is the most likely state, and $\mathrm{erf}(y)$ is the error function.

Using Eq. (\ref{error_prob}), the total probability of error is:

\begin{equation}
\mathrm{P}_{\mathrm{E}} = 1 - \frac{1}{2}\big( 1 + \mathrm{erf}(\sqrt{2}\alpha) \big) = \mathrm{QNL}_{\mathrm{BPSK}}
\end{equation}

For QPSK states $| \alpha_{k} \rangle \in \{| \alpha e^{i k\frac{\pi}{2}} \rangle\}$, where $k \in \{ 0, 1, 2, 3 \}$, the QNL corresponds to the probability of error of an ideal heterodyne measurement \cite{weedbrook12}, which performs a projection onto coherent states and measures both quadratures of the input state simultaneously. The probability of correct discrimination of state $| \alpha_{k} \rangle$ is given by \cite{proakis00}:
\begin{equation}
\mathrm{P}(\alpha_{k} | \alpha_{k}) = \frac{1}{4} \Bigg( 1 + \mathrm{erf} \Big(\frac{\alpha}{\sqrt{2}} \Big) \Bigg)^{2}
\end{equation}

Then, the QNL for QPSK states is \cite{kikuchi16, weedbrook12}:
\begin{equation}
\mathrm{P}_{\mathrm{E}} = 1 - \frac{1}{4} \Bigg( 1 + \mathrm{erf} \Big(\frac{\alpha}{\sqrt{2}} \Big) \Bigg)^{2} = \mathrm{QNL}_{\mathrm{QPSK}}
\end{equation}

While the homodyne measurement is known to be the optimal Gaussian measurement for the discrimination of two coherent states, the ultimate Gaussian limit for coherent multistate discrimination is not known. Therefore, there may be strategies based on Gaussian operations and measurements \cite{weedbrook12} that provide advantages over the heterodyne measurement \cite{wiseman98}.

In a general \emph{M}-PSK encoding $| \alpha_{k} \rangle \in \{| \alpha e^{i k\frac{\pi}{2}} \rangle\}$, where $k \in \{ 0, 1,... M-1 \}$, the probability of correct discrimination can be found through \cite{proakis00}:

\begin{equation}
\mathrm{P}(\alpha_{k} | \alpha_{k}) = \frac{1}{\pi} \iint\limits_{R} e^{-|re^{i\theta} - \alpha_{k}|^{2}} r dr d\theta
\end{equation}

The $\mathrm{QNL}_{\mathrm{MPSK}}$ is then obtained by using Eq. (\ref{error_prob}).

\section{State discrimination strategy}\label{AppStDiscStr}

The phase-tracking method builds on the adaptive measurement strategy for QPSK states with PNR detection described in detail in Ref. \cite{becerra15}. In this strategy, the receiver performs $N=7$ adaptive measurements on the input state $|\alpha_{k}  \rangle\in\{|\alpha\rangle,|i\alpha\rangle,|-\alpha\rangle,|-i\alpha\rangle\}$. In each adaptive measurement $j$ $(j = 1, 2, ..., N)$, the LO is prepared in a state hypothesis $|\beta_{j}\rangle$, and displaces the input state $|\alpha_{k} \rangle$ to $\hat{D}(-\beta_{j})|\alpha_{k} \rangle$. Note that for a correct hypothesis $\beta_{j}=\alpha_{k}$, the input state $|\alpha_{k} \rangle$ is displaced to the vacuum state $|0\rangle$. The displaced state $\hat{D}(-\beta_{j})|\alpha_{k} \rangle$ is then detected with a PNR detector with number resolution $m$, ideally described by operators $\hat{\Pi}_{n} = |n \rangle \langle n|$ for $n = 1, 2, ..., m-1$ and $\hat{\Pi}_{m} = \hat{I} - \sum_{i=0}^{m-1}\hat{\Pi}_{i}$. The strategy uses a maximum $a$ $posteriori$ probability (MAP) criterion and a recursive Bayesian updating \cite{becerra15}. Given a photon number detection $d_{j}$ and the hypothesis $\beta_{j}$  in adaptive measurement $j$, the strategy estimates the posterior Bayesian probabilities for input states and the most likely state. In subsequent adaptive measurements, the LO tests this most likely state, and prior probabilities are updated according to Bayes' theorem. Recursive application of this method during all adaptive measurements results in a final estimate $\theta_{\mathrm{disc}}$ of the possible input state, which corresponds to the most likely state at the end of the last adaptive measurement $N$, $\beta_{N+1}$. This most likely state corresponds to the answer to the state discrimination problem, and the discrimination strategy allows for surpassing the QNL. After a discrimination measurement, the data collected during $N$ adaptive measurements consists of $N$ photon counting detections $\{d_{1},d_{2},...,d_{N}\}$, together with the phases of the most likely states $\beta_{j}$ in each adaptive measurement $\{\theta_{1},\theta_{2},...,\theta_{N}\}$. Assuming that the answer to the state discrimination problem is correct, the phase $\theta_{\mathrm{disc}} = \mathrm{arg}\{\beta_{N+1}\}$ then corresponds to the phase of the input state, so that $\delta_{j}=\theta_{j}-\theta_{\mathrm{disc}}$ are the relative phases of the input state and the LO during each adaptive measurement. The pairs $\{d_{j}, \delta_{j}\}$ correspond to samples of phase space that can be used to estimate the phase offset caused by the channel for performing phase-tracking.

\section{Phase estimator and performance} \label{AppPhEstPerf}

\begin{figure*}[!tbp]
\centering
\includegraphics[width = 0.95\textwidth]{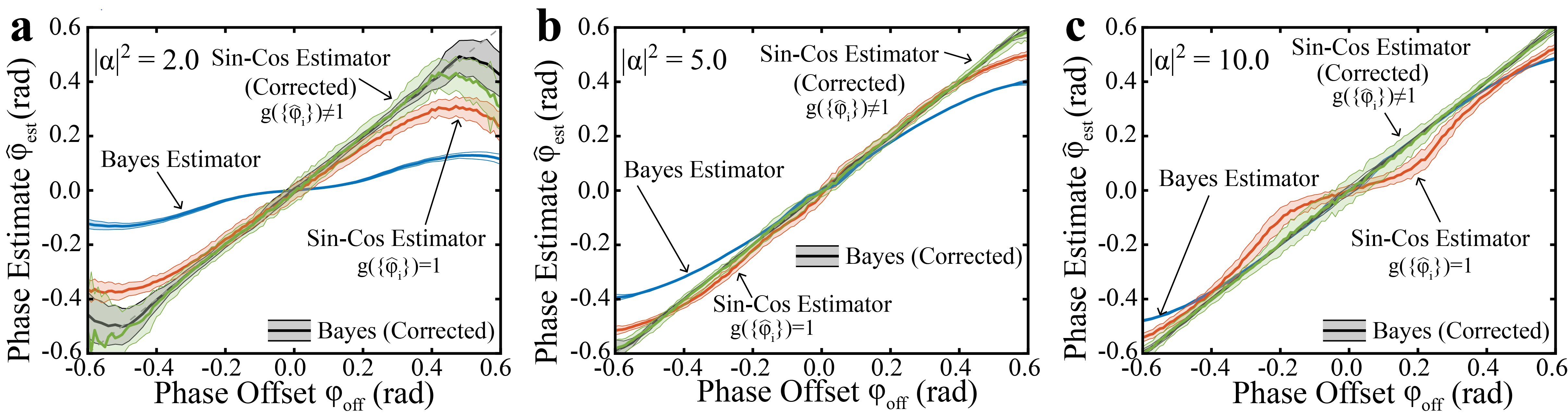}
\caption{ \textbf{Phase estimation and estimator performance.} (a)-(c) Phase estimates $\hat{\phi}$ as a function of applied phase $\phi_{\mathrm{off}}$ for $|\alpha|^{2} = 2.0, 5.0$, and $10.0$, respectively, for the sine-cosine (Sin-Cos) estimator $\hat{\phi}_{\mathrm{est}}$ implemented in our demonstration, and for a Bayesian estimator $\hat{\phi}_{B}$ (blue). The green line shows the corrected $\hat{\phi}_{\mathrm{est}}$ with the optimal gain function $g(\{\hat{\phi}_{i}\})\ne1$. The orange line shows the uncorrected $\hat{\phi}_{\mathrm{est}}$ with $g(\{\hat{\phi}_{i}\})=1$. For the Bayesian estimator, the blue line shows the uncorrected $\hat{\phi}_{B}$, and the black line shows the corrected $\hat{\phi}_{B}$. The solid lines represent the average of 100 Monte Carlo samples and shaded regions correspond to one standard deviation for each value of the applied phase offset $\phi_{\mathrm{off}}$.}
\label{estimator}
\end{figure*}

The method for phase-tracking for non-Gaussian receivers uses the data collected from the state discrimination measurement, consisting of the pairs $\{d_{j}, \delta_{j}\}$, to estimate and correct for the relative phase between the input state and the local oscillator (LO) in real time. This method works in conjunction with the state discrimination strategy and requires no extra resources such as strong phase reference pilot pulses or performing additional measurements for phase estimation. For the adaptive non-Gaussian discrimination measurement in Ref. \cite{becerra15} with photon number resolution (PNR) of 3, PNR(3), the receiver samples four photon number distributions $P_{0}(n_{k}|\delta=0)$, $P_{\pi/2}(n_{k}|\delta=\pi/2)$, $P_{\pi}(n_{k}|\delta=\pi)$, and $P_{3\pi/2}(n_{k}|\delta=3\pi/2)$ for $\delta = \{0, \pi/2, \pi, 3\pi/2\}$. These photon number distributions can be used to obtain different estimators for the phase offset $\phi_{\mathrm{off}}$ (or the applied phase $\phi_{\mathrm{app}}$ in the experimental investigation described in the main manuscript). We note that the photon number distributions $P_{0}(n_{k}|\delta)$ can represent the rows of a $4\times4$ matrix. In general, these distributions can be arranged as rows of a $(\mathrm{PNR}+1) \times M$ detection matrix for $M$-ary shift keyed states and PNR of the measurement. Below, we describe two estimators: one based on the differences of the mean photon numbers $\langle n\rangle_\delta$ of these distributions referred to as ``sine-cosine estimator'' that is implemented in our demonstration, and one that is a Bayesian estimator.

\subsection{\textbf{Sine-cosine estimator}}

The sine-cosine estimator, as implemented in the experimental demonstration described in the main text, uses the differences of the average of detected photon numbers $\langle n\rangle_\delta$ to obtain a final estimate $\hat{\phi}_{\mathrm{est}}$ of the phase offset $\phi_{\mathrm{off}}$ (or the applied phase $\phi_{\mathrm{app}}$ in the actual experiments) based on the collected data from $N_{\textrm{avg}} \times 500$ channel transmissions.

As a first step, the estimator obtains two initial estimates: $\hat{\phi}_{c}$ and $\hat{\phi}_{s}$. {\color{black}These estimates are obtained from the photon number distributions $P_{\delta}(n_{k}|\delta)$ of the observed data from the state discrimination measurement for the relative phases between the input state and LO $\delta = \{0, \pi/2, \pi, 3\pi/2\}$ (see Sec. II of the manuscript).
Under a situation where there is a phase offset $\phi_{\mathrm{off}}$, interference visibility $\xi$, dark count rate $\nu$, and $N=7$, the mean photon numbers of the distributions $P_{\delta}(n_{k}|\delta)$ are:}

\begin{eqnarray}\label{Pmpn}
\langle n\rangle_{0}=  2\eta\frac{|\alpha|^2}{N}[1-\xi\textrm{cos}(\phi_{\mathrm{off}})]+\nu,
 \\ \nonumber
\langle n\rangle_{\pi/2}=   2\eta\frac{|\alpha|^2}{N}[1-\xi\textrm{sin}(\phi_{\mathrm{off}})]+\nu,
 \\ \nonumber
 \langle n\rangle_{\pi}=  2\eta\frac{|\alpha|^2}{N}[1+\xi\textrm{cos}(\phi_{\mathrm{off}})]+\nu,
 \\ \nonumber
\langle n\rangle_{3\pi/2}=   2\eta\frac{|\alpha|^2}{N}[1+\xi\textrm{sin}(\phi_{\mathrm{off}f})]+\nu.
 \nonumber
 \end{eqnarray}

{\color{black}Combining the equations for $\langle n\rangle_{0}$ and $\langle n\rangle_{\pi}$ in Eq. (\ref{Pmpn}) we can obtain samples for the quantity $\textrm{cos}(\phi_{\mathrm{off}})$ in terms of $\langle n\rangle$, $\eta$, $|\alpha|^2$, $\xi$, $\nu$, and $N$. In a similar way, samples for $\textrm{sin}(\phi_{\mathrm{off}})$ can be obtained from the equations for $\langle n\rangle_{\pi/2}$ and $\langle n\rangle_{3\pi/2}$. These samples can be used to obtain an estimate of the expected values from the average over $500$ channel transmissions. For a large number of data samples, we expect that the average of $\textrm{cos}(\phi_{c})$ over these channel transmissions approach the cosine of the average $\bar{\phi}_{\mathrm{off}}$ of the actual phase offset $\phi_{\mathrm{off}}$ over these channel transmissions, $\textrm{cos}(\bar{\phi}_{\mathrm{off}})$.
%, and $\textrm{sin}(\bar{\phi}_{c})$, where $\bar{\phi_{c,s}}$
%are the average of $phi_{c,s}$
We define this average $\bar{\phi}_{\mathrm{off}}$ from the cosine function as the estimate $\hat{\phi}_{c}$. Similarly, the estimate $\hat{\phi}_{s}$ is obtained from the samples of $\textrm{sin}(\phi_{\mathrm{off}})$.
These estimates $\hat{\phi}_{c}$ and $\hat{\phi}_{s}$ can be expressed in terms of the estimates of the mean photon numbers $\langle n\rangle_\delta$ for $500$ channel transmissions:

\begin{align}\label{Phi_c}
 \langle n \rangle_{\pi} - \langle n \rangle_{0} &= C(|\alpha|^{2}) \textrm{cos}(\hat{\phi}_{c}),
 \\ \nonumber
 \hat{\phi}_{c} &= \textrm{arccos}\bigg[\frac{\langle n \rangle_{\pi} - \langle n \rangle_{0}}{ C(|\alpha|^{2})}\bigg]
 \nonumber
 \end{align}
 and
\begin{align} \label{Phi_s}
\langle n \rangle_{3\pi/2} - \langle n \rangle_{\pi/2} &= C(|\alpha|^{2}) \textrm{sin}(\hat{\phi}_{s}),
  \\ \nonumber
 \hat{\phi}_{s} &= \textrm{arcsin}\bigg[\frac{\langle n \rangle_{3\pi/2} - \langle n \rangle_{\pi/2}}{ C(|\alpha|^{2})}\bigg]
 \nonumber
  \end{align}
 with
\begin{equation}
 C(|\alpha|^{2}) = f(|\alpha|^{2}) \times 4|\alpha|^{2}\eta \xi/N
\label{C_alpha}
\end{equation}

where $f(|\alpha|^{2})$ is a factor arising from non-zero probability of error of the discrimination strategy. Here, $N=7$ is the number of adaptive measurements, $\eta$ is the detection efficiency, and $\xi$ is visibility of the displacement operation by interference. As a second step, the two initial estimates $(\hat{\phi}_{c}, \hat{\phi}_{s})$ are combined in a weighted average to form a phase estimate $\hat{\phi}_{i}$ every 500 pulses:

\begin{equation}
\hat{\phi}_{i}=\mathrm{sign}(\hat{\phi}_{s})\frac{|\hat{\phi}_{i}|+r(|\alpha|^{2})|\hat{\phi}_{s}|}{1+r(|\alpha|^{2})}.
\end{equation}

The weight factor $r(|\alpha|^{2})$ is used to increase the linearity of the final estimate $\hat{\phi}_{\mathrm{est}}$ as a function of $\phi_{\mathrm{off}}$, while reducing its variance near the edges of the capture range $R$ in our experiment $R=\pm 0.6$ rad. As a final step, the final estimate $\hat{\phi}_{\mathrm{est}}$ is obtained from the average of $N_{\textrm{avg}}$ estimates $\hat{\phi}_{i}$ with a gain factor $g({\hat{\phi}_{i}})$

\begin{equation}
\hat{\phi}_{\mathrm{est}}=g(\langle\hat{\phi}\rangle)\frac{1}{N_{\textrm{avg}}}\sum \hat{\phi}_{i}=g(\langle\hat{\phi}\rangle)\langle\hat{\phi_{i}}\rangle.
\label{Phi_est}
\end{equation}

The gain factor $g(\langle\hat{\phi}\rangle)$ depends on the average of the $N_{\textrm{avg}}$ estimates $\{\hat{\phi}\}=\{\hat{\phi}_{1}, \hat{\phi}_{2}, ..., \hat{\phi}_{N_{\textrm{avg}}}\}$ and is used to further increase the linearity with respect to the actual phase offset $\phi_{\mathrm{off}}$, as described below.

To obtain the final phase estimate $\hat{\phi}_{\mathrm{est}}$ of $\phi_{\mathrm{off}}$, the estimator aims to find the optimal values for the factors $r(|\alpha|^{2})$, $f(|\alpha|^{2})$, and $g(\langle\hat{\phi}_{i}\rangle)$, which depend on the input mean photon number $|\alpha|^{2}$, the estimates $\hat{\phi}_{i}$, and the experimental detection efficiency $\eta$, visibility $\xi$, and dark counts. We find the optimal values of $r(|\alpha|^{2})$, $f(|\alpha|^{2})$, and $g(\langle\hat{\phi}_{i}\rangle)$ using numerical approaches based on Monte Carlo simulations of the experiment with the following steps:
\\\\
\textbf{1}.- Find the optimal value of $r(|\alpha|^{2})$ ($r_{opt}(|\alpha|^{2})$) that minimizes the difference $|\hat{\phi}_{i}-\phi_{\mathrm{off}}|$ at the extreme points of the capture range $R=\pm0.6$ rad.
\\
\textbf{2}.- Given $r_{opt}(|\alpha|^{2})$, find the optimal value of $f(|\alpha|^{2})$ [$f_{opt}(|\alpha|^{2})$] by minimizing the $\chi^2$ between the estimated phase $\hat{\phi}_{i}$ and the actual phase offset $\hat{\phi}_{\mathrm{off}}$.
\\
\textbf{3}.- Given $r_{opt}(|\alpha|^{2})$ and $f_{opt}(|\alpha|^{2})$, find the gain parameter $g(\langle\hat{\phi}\rangle)$ that makes the final estimate $\hat{\phi}_{\mathrm{est}}$ as linear as possible with respect to the applied phase offset $\phi_{\mathrm{off}}$.
\\\\
This procedure yields the final estimate $\hat{\phi}_{\mathrm{est}}$, and the optimal parameters $r_{opt}(|\alpha|^{2})$, $f_{opt}(|\alpha|^{2})$, and $g(\langle\hat{\phi}\rangle)$ from Monte Carlo simulations that we use for the experimental demonstration of phase-tracking for non-Gaussian receivers.

Figure (\ref{estimator}) shows the final estimate $\hat{\phi}_{\mathrm{est}}$ for mean photon numbers $|\alpha|^{2}$=2.0, $|\alpha|^{2}$=5.0, and $|\alpha|^{2}$=10.0 with optimized gain parameter $g(\langle\hat{\phi}\rangle)\neq1$ (green) and with $g(\langle\hat{\phi}\rangle)=1$ (blue). Note that the estimate $\hat{\phi}_{\mathrm{est}}$ with $g(\langle\hat{\phi}\rangle)\neq1$ shows a closer linear relation with $\phi_{\mathrm{off}}$ compared to the case with $g(\langle\hat{\phi}\rangle)=1$. Below we describe the procedure to obtain $r_{opt}(|\alpha|^{2})$, $f_{opt}(|\alpha|^{2})$, and $g(\langle\hat{\phi}\rangle)$.

This procedure reduces the bias of the final estimate $\hat{\phi}_{\mathrm{est}}$ with respect to the true value of the phase offset $\phi_{\mathrm{off}}$. The estimates $\hat{\phi}_{c}$ and $\hat{\phi}_{s}$ are initially biased. $\hat{\phi}_{s}$ is biased towards zero phase as the magnitude of $\phi_{\mathrm{off}}$ increases. $\hat{\phi}_{c}$ cannot provide information about the sign of the phase offset $\phi_{\mathrm{off}}$, and biases the estimates towards positive values. These biases result in a bias of the combined estimator $\phi_{i}$, and $r(|\alpha|^{2})$ aims to reduce this bias. The final estimate $\hat{\phi}_{\mathrm{est}}$ is also biased for large values of the phase offsets, as can be seen in Fig. (\ref{estimator}), but unbiased for phase offsets near zero. The optimization of the gain function $g(\langle\hat{\phi}\rangle)$ allows for minimizing the overall bias of $\hat{\phi}_{\mathrm{est}}$, which can be mostly suppressed for $|\alpha|^{2} \geq 5$.}=
\\
\\
\textbf{Step 1.- Optimal value of $r(|\alpha|^{2})$}
\\
The parameter $r(|\alpha|^{2})$ is a weight factor for the contributions of the initial estimates $\hat{\phi}_{c}$ and $\hat{\phi}_{s}$ to the estimate $\hat{\phi}_{i}$, and its optimal value is chosen to reduce the variance of the final estimate $\hat{\phi}_{\mathrm{est}}$ near the end points of the experimental capture range range $R=\pm 0.6$ rad. The collected data $\{d_{j}, \delta_{j}\}$ during the discrimination measurement provides samples for phase estimation which are mostly $\delta_{j}=0$. This is because the discrimination strategy is based on hypothesis testing by displacements to the vacuum state \cite{becerra15}, and the displaced state spends most of the time in the vacuum state. This means the distribution $P_{0}(n_{k}|\delta=0)$ is populated at a higher rate than $P_{\pi/2}$, $P_{\pi}$, and $P_{3\pi/2}$, and provides more data for the initial estimate $\hat{\phi}_{c}$ compared to $\hat{\phi}_{s}$ [see Eqs. (\ref{Phi_c}) and (\ref{Phi_s})]. As a result, $\hat{\phi}_{c}$ gives a much better estimate with smaller variance. However, $\hat{\phi}_{c}$ does not give any sign information about the applied phase $\phi_{\mathrm{off}}$, and is less sensitive to small phase offsets around $\phi_{\mathrm{off}}=0$. On the contrary, $\hat{\phi}_{s}$ is more sensitive to small phase offsets and contains the sign information of $\phi_{\mathrm{off}}$, but is a worse estimate with a much larger variance. The optimal value of $r(|\alpha|^{2})$ seeks to balance the contribution of these two initial estimates to minimize $|\hat{\phi}_{i}-\phi_{\mathrm{off}}|$ at $R=\pm 0.6$ rad. This optimization has the overall effect of reducing the variance of the final estimate $\hat{\phi}_{\mathrm{est}}$ at these points, where this estimator shows the greatest variance.

To find the optimal value of $r(|\alpha|^{2})$ [$r_{\mathrm{opt}}(|\alpha|^{2})$] we fix the mean photon number $|\alpha|^2$ and $N_{\textrm{avg}}$. We use Monte Carlo simulations to obtain the weighted average $\hat{\phi}_{i}$ with $f(|\alpha|)=1$ as a function of $\phi_{\mathrm{off}}$ for different values of $r(|\alpha|^{2})$. We then obtain a final average for the phase estimate $\langle\hat{\phi_{i}}\rangle=\sum\hat{\phi}_{i}/N_{\textrm{avg}}$ after $N_{\textrm{avg}}$ realizations for applied phases $\phi_{\mathrm{off}}$ within the range $R=\pm0.6$ rad. We observe that $\langle\hat{\phi_{i}}\rangle$ is in general a non-linear function of $\phi_{\mathrm{off}}$. The optimal $r_{\mathrm{opt}}(|\alpha|^{2})$ is obtained by finding the value of $r(|\alpha|^{2})$ that minimizes the average of $|\langle\hat{\phi}_{i}\rangle-\phi_{\mathrm{off}}|$ at $\phi_{\mathrm{off}}=\pm0.6$ rad. This condition increases the linearity of the final phase estimate $\hat{\phi}_{\mathrm{est}}$ with respect to the applied phase $\phi_{\mathrm{off}}$ and reduces its variance.

We note that $\hat{\phi}_{\mathrm{est}}$ is obtained by multiplying $\langle\hat{\phi_{i}}\rangle$ by a gain function $g(\langle\hat{\phi_{i}}\rangle)$. As described in Step 3, $g(\langle\hat{\phi_{i}}\rangle)$ is obtained by inverting the relation of $\langle\hat{\phi_{i}}\rangle$ as a function of $\phi_{\mathrm{off}}$. Minimizing $|\langle\hat{\phi}_{i}\rangle-\phi_{\mathrm{off}}|$ at $\phi_{\mathrm{off}}=\pm0.6$ rad prevents a large value of the gain $g(\langle\hat{\phi_{i}}\rangle)$ at these points, which would result in a great increase in the variance of the final estimate $\hat{\phi}_{\mathrm{est}}$. Therefore, determining $r_{\mathrm{opt}}(|\alpha|^{2})$ that minimizes $|\hat{\phi}_{i}-\phi_{\mathrm{off}}|$ at $(\pm0.6)$ rad, results in a gain $g(\langle\hat{\phi_{i}}\rangle)\approx1$, reducing the variance of the final estimate $\hat{\phi}_{\mathrm{est}}$.
\\
\\
\textbf{Step 2.- Optimal value of $f(|\alpha|^{2})$}
\\
The parameter $f(|\alpha|^{2})$ in Eq. (\ref{C_alpha}) aims to reduce the effect of the non-zero probability of error $P_\textrm{E}$ in the state discrimination measurement. Any state discrimination error causes errors when populating the photon number distributions $P_{\delta}(n_{k}|\delta)$, $\delta = \{0, \pi/2, \pi, 3\pi/2\}$. In a situation with no errors, $P_\textrm{E}=0$, $f(|\alpha|^{2})= 1$ and on average $C(|\alpha|^2)=4|\alpha|^{2}\eta \xi/N$. However, when $P_\textrm{E}\ne0$ the distributions $P_{\delta}(n_{k}|\delta)$ are some times populated incorrectly, which produces changes in their mean photon numbers $\langle n\rangle_{\delta}$ in Eq. (\ref{Phi_c}) and (\ref{Phi_s}). This makes $C(|\alpha|^2)\neq4|\alpha|^{2}\eta \xi/N$, and $f(|\alpha|^{2})\neq1$.

We note that the distribution $P_{\pi}(n_{k}|\delta=\pi)$ can only be incorrectly populated with samples from the other three distributions $P_{0}$, $P_{\pi/2}$, and $P_{3\pi/2}$, which have smaller mean photon numbers. As a result, any discrimination error causes the estimated $\langle n \rangle_{\pi}$ to be smaller than the true value on average.
Similarly, the distribution $P_{0}(n_{k}|\delta=0)$ can only be incorrectly populated with samples from distributions $P_{\pi/2}$, $P_{\pi}$, and $P_{3\pi/2}$ with larger mean photon numbers. Then, $P_\textrm{E}\ne0$ causes the estimated $\langle n \rangle_{0}$ to increase on average.
The overall effect of having $P_\textrm{E}\ne0$ is to reduce the difference $\langle n \rangle_{\pi}-\langle n \rangle_{0}$, such that $\langle n \rangle_{\pi}-\langle n \rangle_{0}<4|\alpha|^{2}\eta \xi/N$.
As a result, the estimate $\hat{\phi}_{c}$ in Eq. (\ref{Phi_c}) with $f(|\alpha|^{2})=1$ will have a non-zero value when $\phi_{\mathrm{off}}=0$ that depends on the probability of error $P_\textrm{E}$.

The effect of having $P_\textrm{E}\ne0$ can be reduced by finding the value of the parameter $f(|\alpha|^2)$ that makes the estimates $\hat{\phi}_{c}$ and $\langle\hat{\phi_{i}}\rangle$ close to zero when $\phi_{\mathrm{off}}=0$. The procedure to find the optimal value $f_{\mathrm{opt}}(|\alpha|^{2})$ consists of using Monte Carlo simulations for different values of $f(|\alpha|^{2})$ with $r_{\mathrm{opt}}(|\alpha|^{2})$ found in Step 1. The optimal value $f_{\mathrm{opt}}(|\alpha|^{2})$ is the one that minimizes the $\chi^2$ between $\langle\hat{\phi_{i}}\rangle$ and $\phi_{\mathrm{off}}$, where a linear dependence is expected. To verify this optimal value we use $\approx 10^{6}$ Monte Carlo runs with $\phi_{\mathrm{off}}=0$ to obtain the expected difference $\mathrm{E} [ \langle n \rangle_{\pi}-\langle n \rangle_{0}  | \phi_{\mathrm{off}ff}=0 ]$. The value $f_{\mathrm{opt}}(|\alpha|^{2})$ should then be approximately:

\begin{equation}
f_{\mathrm{opt}}(|\alpha|^{2}) \approx \frac{ \mathrm{E} [ \langle n \rangle_{\pi}-\langle n \rangle_{0} | \phi_{\mathrm{off}}=0 ]}{4|\alpha|^{2}\eta \xi/N}
\end{equation}

 Table (S1) shows examples of the optimal values of $r(|\alpha|^{2})$ and $f(|\alpha|^2)$ for different mean photon numbers of the input state $|\alpha|^{2}$ and $N_{\textrm{avg}}$.
\\
\begin {table}
\begin{center}
\caption{Optimal values $f_{\mathrm{opt}}(|\alpha|^2)$ and $r_{\mathrm{opt}}(|\alpha|^{2})$ for $|\alpha|^2=2.0$, 5.0, and 10.0 with $N_{\textrm{avg}}$=40, 20, and 4, respectively.}
\begin{tabular}{ |c|c|c|c| }
 \hline
$|\alpha|^{2}$ & $N_{\textrm{avg}}$ & $f_{\mathrm{opt}}(|\alpha|^2)$ & $r_{\mathrm{opt}}(|\alpha|^{2})$ \\
 \hline
 2.0 & 40 & 1.210 & 0.250 \\
 5.0 & 20 & 0.895 & 0.333\\
 10.0 & 4 & 0.650 & 0.250\\
 \hline
\end{tabular}
\end{center}
\end {table}
\\
\textbf{Step 3.- Gain function $g(\langle\hat{\phi_{i}}\rangle)$}
\\
The final estimate $\hat{\phi}_{\mathrm{est}}$ is obtained from the product of the average of $N_{\textrm{avg}}$ estimates $\hat{\phi}_{i}$ and a gain factor $g(\langle\hat{\phi_{i}}\rangle)$, as shown in Eq. (\ref{Phi_est}). The gain function $g(\langle\hat{\phi_{i}}\rangle)$ is solely a function of the average phase estimate $\langle{\hat{\phi}}_{i}\rangle$ , and ideally maps $\langle{\hat{\phi}}_{i}\rangle$ onto the phase offset $\phi_{\mathrm{off}}$ with a linear dependence with unit slope. We obtain the gain function $g(\langle\hat{\phi_{i}}\rangle)$ by using Monte Carlo simulations with the optimal values of $r_{\mathrm{opt}}(|\alpha|^{2})$ and $f_{\mathrm{opt}}(|\alpha|^{2})$ to obtain the dependence of the quantity $(\phi_{\mathrm{off}} / \langle\hat{\phi_{i}}\rangle)$ as a function of $\langle \hat{\phi_{i}} \rangle$. The quantity $(\phi_{\mathrm{off}} / \langle \hat{\phi}_{i}\rangle)$ shows in general a nonlinear dependence with $\langle \hat{\phi_{i}} \rangle$, and this dependence corresponds to $g(\langle\hat{\phi_{i}}\rangle)$. We fit the quantity $(\phi_{\mathrm{off}} / \langle \hat{\phi}_{i}\rangle)$ using a smoothing spline, and this spline is defined as the gain function $g(\langle\hat{\phi_{i}}\rangle)$.

The gain function $g(\langle\hat{\phi_{i}}\rangle)$ obtained in this way allows to linearize the phase estimator with respect to known applied phase offsets $\phi_{\mathrm{off}}$. Figure \ref{estimator} shows $\hat{\phi}_{\mathrm{est}}$ for input mean photon numbers $|\alpha|^{2} = 2.0,$ $5.0$, and $10.0$ with optimal values of $g(\langle\hat{\phi_{i}}\rangle)$ $(\ne1)$ in green; and with $g(\langle\hat{\phi_{i}}\rangle)=1$ in orange. In our experimental demonstration, the method for phase-tracking is set to generate estimates every $500\times N_{\textrm{avg}}$ transmissions through the channel, and subsequently applies a phase correction to the local oscillator every 500$\times N_{\mathrm{avg}}$ shots of the experiment, allowing to perform phase-tracking and phase correction at a rate of $f_{\mathrm{PT}} \approx 23/N_{\mathrm{avg}}$ Hz.

\subsection{\textbf{Bayesian estimator}}

\begin{figure*}
\centering\includegraphics[width = 0.95\textwidth]{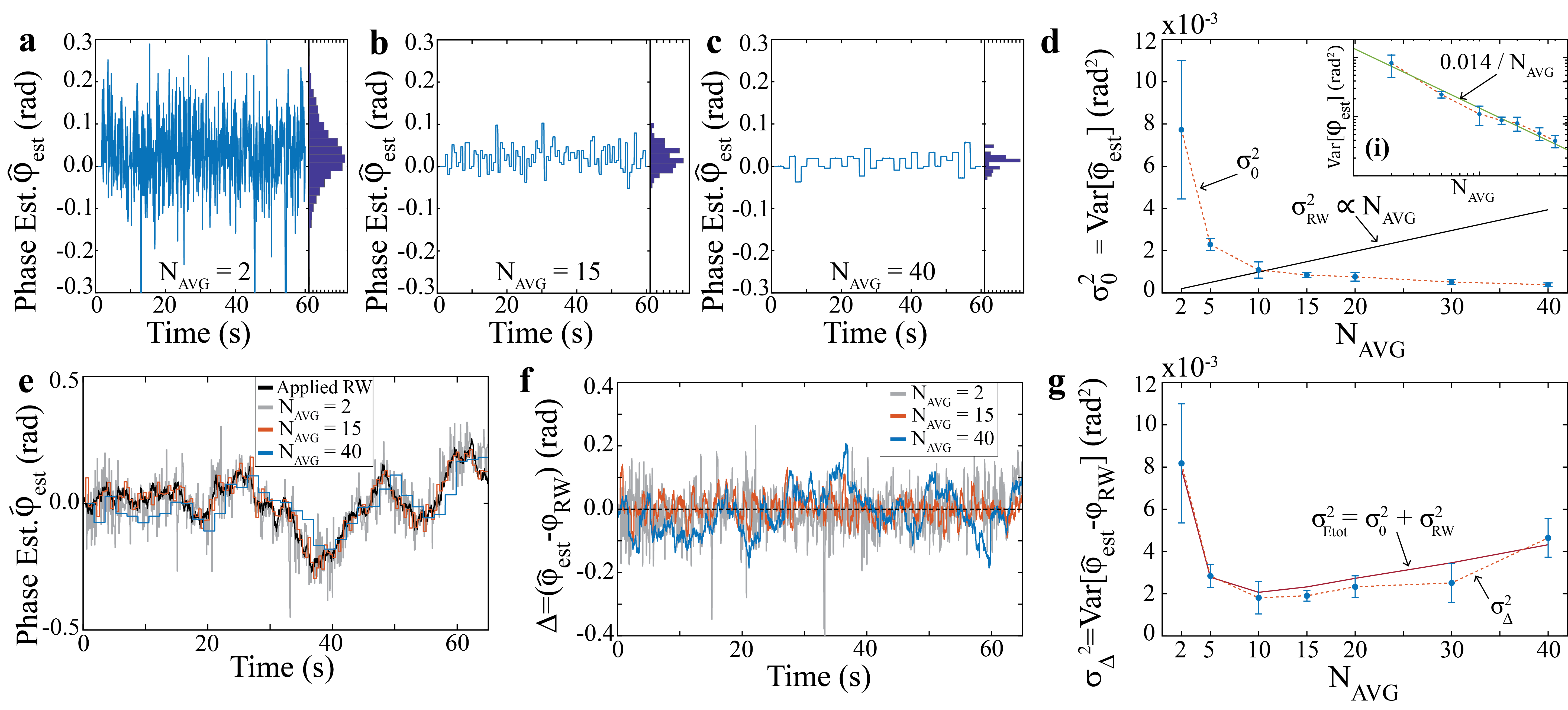}
\caption{\textbf{Estimator variance of the phase-tracking method.} Experimentally obtained variance of the estimates $\hat{\phi}_{\mathrm{est}}$ without ($\sigma^2_{0}$) and with ($\textrm{Var}[\hat{\phi}_{\mathrm{est}}-\phi_{\mathrm{RW}}]$) added Gaussian phase noise for $|\alpha|^{2} = 5.0$ for different $N_{\textrm{avg}}$ in the algorithm for phase-tracking. (a-c) phase-tracking estimates $\hat{\phi}_{\mathrm{est}}$ with zero applied phase ($\phi_{\mathrm{app}}= 0$) for (a) $N_{\textrm{avg}}$=2, (b) $N_{\textrm{avg}}$=15, and (c) $N_{\textrm{avg}}$=40 as a function of time from $t$=0 to 60 s. (d) Variance $\sigma^2_{0}$ over 60 s with $\phi_{\mathrm{app}}= 0$ (blue points) as a function of $N_{\textrm{avg}}$. Error bars represent one standard deviation over 5 different experimental runs. The black solid line shows the expected accumulated variance of random walks $\sigma^2_{\mathrm{RW}}$ in phase with $\sigma_{1}=5$ mrad, similar to Fig. 4(e) in the main manuscript, for different estimators with different $N_{\textrm{avg}}$. Inset (\emph{i}) shows $\sigma^2_{0}$ on a log-log scale with a linear fit indicating a $1/N_{\mathrm{avg}}$ scaling, as expected for statistical uncertainties. (e) Estimated phase for a single applied random walk ($\phi_{\mathrm{app}}=\phi_{\mathrm{RW}}$) (black line) for three estimators with $N_{\textrm{avg}}$=2, 15, and 40, in grey, orange, and blue, respectively. (f) Difference between estimated and applied phases $\Delta=\hat{\phi}_{\mathrm{est}}-\phi_{\mathrm{RW}}$ in (e). (g) Variance of the total estimates $\sigma^2_{\Delta}$ for the difference $\Delta$ in (f) as a function of $N_{\textrm{avg}}$. Error bars represent one standard deviation from 5 different experimental runs. The red solid line shows the expected total variance $\sigma^2_{\mathrm{Etot}}=\sigma^2_{0}+\sigma^2_{\mathrm{RW}}$ which is composed of $\sigma^2_{0}$ and $\sigma^2_{\mathrm{RW}}$ from (d), showing good agreement with the observed $\sigma^2_{\Delta}$ in the experiment. Note that there is an optimal operation point at $N_{\textrm{avg}}\approx10$, where the total variance $\sigma^2_{\Delta}$ is minimum.
\label{var}
}
\end{figure*}
A second possible estimator of $\phi_{\mathrm{off}}$ based on the collected data $\{d_{j}, \delta_{j}\}$ from the state discrimination measurement is the Bayesian estimator. For a Bayes estimator, the photon number distributions $P_{\delta}(n_{k}|\delta)=P(n)$ are converted into distributions over the phase $P(\phi | n)$ through Bayes' theorem
\begin{equation}
P(\phi | n)P(n) = \mathcal{L}(n | \phi)P(\phi),
\end{equation}
where $\mathcal{L}(n | \phi)$ are likelihood functions and $P(\phi)$ a prior phase distribution. Given the collected data from the state discrimination measurement $\{\mathrm{data}\}=\{d_{j}, \delta_{j}\}$ and assuming some prior distribution $P(\phi)$, a phase estimate can be obtained by forming the posterior probability distribution over phase given by:

\begin{equation}
P(\phi|\{\mathrm{data}\}) = \mathcal{N}P(\phi)\prod_{n=0}^{3}\prod_{m=0}^{3}\mathcal{L}(n |\phi - \delta_{m})^{N_{n,m}}
\end{equation}

where $\mathcal{N}$ is a normalization factor and $N_{n,m}$ is the number of times that $n$ photons were detected given that $\delta_{m} = \theta_{m}-\theta_{\mathrm{disc}} = m\pi/2$. Here $\theta_{\mathrm{disc}}$ is the answer to the state discrimination problem about the input state. The values of $N_{n,m}$ correspond to elements of the matrix of the photon detections for given $\{\delta_{m}\}$.

The likelihood function $\mathcal{L}(n |\phi - \delta_{m})$ is given by:
\begin{align}
&\mathcal{L}(n |\phi - \delta_{m}) = \frac{\langle n\rangle^{n}}{n!} e^{-\langle n \rangle}
\nonumber
\end{align}
with
\begin{align}
&\langle n \rangle = 2\eta|\alpha|^2\big [1 - \xi \mathrm{cos}(\phi - \delta_{m})\big] + \nu
\end{align}
where $\eta$, $\xi$, and $\nu$ are the detection efficiency, interference visibility, and dark counts, respectively. The phase estimate $\hat{\phi}_{B}$ for the Bayesian estimator is then given by:
\begin{equation}
\hat{\phi}_{B} = \mathrm{arg}\bigg(\int_{-\pi}^{\pi}e^{i\phi} P(\phi | \{\mathrm{data}\}) d\phi\bigg).
\end{equation}

The Bayesian estimate $\hat{\phi}_{B}$ provides a more precise estimate of the phase offset $\phi_{\mathrm{off}}$ with smaller variance than the sine-cosine estimate $\hat{\phi}_{\mathrm{est}}$, as shown in Fig. \ref{estimator}. However, this estimator is far more computationally difficult to implement experimentally in real time. While in our current experimental setup such a complex estimator cannot be implemented, the sine-cosine estimator described above produces similar results for estimating $\phi_{\mathrm{off}}$ while remaining computationally inexpensive, and allows for real-time estimation and implementation of the phase-tracking method.

\section{Estimator performance as a function of $N_{\textrm{avg}}$} \label{AppEstPerf-Nave}

The performance of the phase-tracking method critically depends on the variance of the estimates $\hat{\phi}_{\mathrm{est}}$. In general, increasing the number of samples $N_{\textrm{avg}}$ to obtain an estimate of the applied phase $\phi_{\mathrm{app}}$ \footnote{We denote $\phi_{\mathrm{app}}$ phases that are actually applied in the experimental studies, and we denote $\phi_{\mathrm{off}}$ phases that are used to find theoretical values for the optical parameters $r_{\mathrm{opt}}(|\alpha|^{2})$, $f_{\mathrm{opt}}(|\alpha|^2)$, and $g(\langle\phi\rangle)$ for phase-tracking.}
improves (reduces) the variance of the estimates $\hat{\phi}_{\mathrm{est}}$. However, increasing $N_{\textrm{avg}}$ also increases the time required to obtain such estimate, thus reducing the phase-tracking bandwidth $f_{\mathrm{PT}}$. In situations with random Gaussian phase noise with bandwidth $f_{\mathrm{RW}}>f_{\mathrm{PT}}$, this reduction in $f_{\mathrm{PT}}$ can significantly increase probability of error $P_{\textrm{E}}$ in the state discrimination measurement, and produce estimates that are far less accurate. As a result, there is a trade-off in the performance of the estimator as a function of $N_{\textrm{avg}}$. While larger values of $N_{\textrm{avg}}$ provide better estimates for constant phases, in situations where the phase is not constant these estimates may not be accurate, limiting the performance of the phase-tracking method. To investigate this trade-off, we experimentally study the estimator variance as a function of $N_{\textrm{avg}}$ in situations with Gaussian-distributed random phase noise.

Figures \ref{var}(a)--(c) shows the estimates from the experiment using the phase-tracking method as a function of time from $t=0$ to $60$ s with zero applied phase ($\phi_{\mathrm{app}}=0$) for $|\alpha|^{2} = 5.0$, for cases with (a) $N_{\textrm{avg}}$=2, (b) $N_{\textrm{avg}}$=15, and (c) $N_{\textrm{avg}}$ = 40, and the corresponding histograms of estimates. We observe that while smaller $N_{\textrm{avg}}$ increases phase-tracking bandwidth $f_{\mathrm{PT}}$, the variance of the estimates for $\phi_{\mathrm{app}}=0$, denoted as $\sigma_{0}^2$, also increases. On the other hand, larger $N_{\textrm{avg}}$ reduces $f_{\mathrm{PT}}$, but also reduces $\sigma_{0}^2$. Figure \ref{var}(d) shows the variance $\sigma_{0}^2$ for estimators with different $N_{\textrm{avg}}$ (blue points) for five experimental runs with $\phi_{\mathrm{app}}=0$. The inset (i) shows the variance $\sigma_{0}^2$ on a log-log scale with a best-fit line showing good agreement with a $1/N_{\textrm{avg}}$ scaling, which is consistent with the statistical uncertainty for a process with random noise.

For situations with random Gaussian phase noise $\phi_{\mathrm{RW}}$ with variance $\sigma^2_{\mathrm{RW}}$, the total variance of the estimates will contain contributions from $\sigma_{0}$ and $\sigma_{\mathrm{RW}}$. The solid black line in Fig. \ref{var}(d) shows the expected accumulated variance $\sigma^2_{\mathrm{RW}}$ for Gaussian random walks between the times to obtain phase estimates, which is proportional to $N_{\textrm{avg}}$. Then, the total expected variance $\sigma_{\mathrm{Etot}}^2$ for situations with Gaussian phase noise will be approximately the sum in quadrature of $\sigma_{0}$ (related to the variance of the estimator with different $N_{\textrm{avg}}$) and $\sigma_{\mathrm{RW}}$ \footnote{We note that the estimator $\hat{\phi}_{\mathrm{est}}$ uses the data from the state discrimination measurement of four coherent states to efficiently estimate a phase offset $\phi_{\mathrm{off}}$. This task is different from the problem of estimating an unknown phase of a known coherent state probe, for which the fundamental lower bound on the variance is known \cite{giovannetti11}, and the variance of $\hat{\phi}_{\mathrm{est}}$ is higher than the fundamental lower bound for that problem \cite{giovannetti11}.}.

Figure \ref{var}(e) shows the performance of the estimator for a given applied Gaussian random walk in phase (black line) for $N_{\textrm{avg}}$=2, 15, and 40, with $\sigma_{1}=5$ mrad ($\sigma_{1}$ as defined in the main text) from $t=0$ to $60$ s. Figure \ref{var}(f) shows the difference between estimated $\hat{\phi}_{\mathrm{est}}$ and the applied phase $\phi_{\mathrm{RW}}$, $\Delta=\hat{\phi}_{\mathrm{est}}-\phi_{\mathrm{RW}}$ from Fig. \ref{var}(e). The variance of $\Delta$ will now contain two contributions: one from the estimator with different $N_{\textrm{avg}}$, ideally given by $\sigma_{0}^2$, and one due to the random walks in phase with variance $\sigma^2_{\mathrm{RW}}$. For $N_{\textrm{avg}}=2$ we expect a large variance of the estimates $\hat{\phi}_{\mathrm{est}}$, as shown in Fig. \ref{var}(a), and due to the relatively high tracking bandwidth $f_{\mathrm{PT}}$, the effect of the random walks is relatively small. On the other hand, for $N_{\textrm{avg}}=40$, the variance of estimates $\hat{\phi}_{\mathrm{est}}$ is expected to be small, as seen in Fig. \ref{var}(c), but due to the low $f_{\mathrm{PT}}$ relative to $f_{\mathrm{RW}}$ ($f_{\mathrm{RW}} = 100\mathrm{Hz}$), the accumulated variance from random walks becomes dominant resulting in larger deviations $\Delta$.

Figure \ref{var}(g) shows the total variance $\sigma^{2}_{\Delta}$ of $\Delta=\phi_{\mathrm{est}}-\phi_{\mathrm{RW}}$ as a function of $N_{\textrm{avg}}$. Error bars represent one standard deviation over five different random walks. The red solid line shows the expected total variance $\sigma^2_{\mathrm{Etot}}=\sigma^2_{0}+\sigma^2_{\mathrm{RW}}$, which contains the contributions from the estimator $\sigma^2_{0}$ and the applied random walks $\sigma_{\mathrm{RW}}^2$. The good agreement between $\sigma^2_{\Delta}$ and $\sigma^2_{Etot}$ indicates that drifts in the experiment are not significant. We observe that there is an optimal value for $N_{\textrm{avg}}\approx10$ that minimizes the total variance $\sigma^2_{\Delta}$. This optimal value of $N_{\textrm{avg}}$ provides a good phase estimate $\hat{\phi}_{\mathrm{est}}$ by increasing the number of samples for parameter estimation, while reducing the effects of errors in state discrimination caused by drifts in phase due to the random walks. We note that optimal values for $N_{\textrm{avg}}$ are larger for smaller values of $f_{\mathrm{RW}}$, and vice-versa.

\section{phase-tracking with different noise bandwidths} \label{AppEstPerf-BW}
Methods for phase-tracking should be able to track phase noise with different bandwidths. In general, the performance of any phase-tracking method to track fast noise depends on how fast reliable phase estimates can be generated and how fast correction can be applied \cite{ip07}, which defines the phase-tracking bandwidth $f_{\mathrm{PT}}$. We investigate the performance of the phase-tracking method for the sub-QNL non-Gaussian receiver to track time-varying random phase noise with different bandwidths $f_{\mathrm{RW}}$ for $|\alpha|^{2} = 5.0$, while keeping the same phase-tracking bandwidth $f_{\mathrm{PT}} = f_{exp}/(500 N_{\mathrm{avg}})\approx1.2$ Hz, with $f_{exp}\approx12$ kHz and $N_{\textrm{avg}} = 20$.

\begin{figure}[!tbp]
\centering\includegraphics[width = 7.5cm]{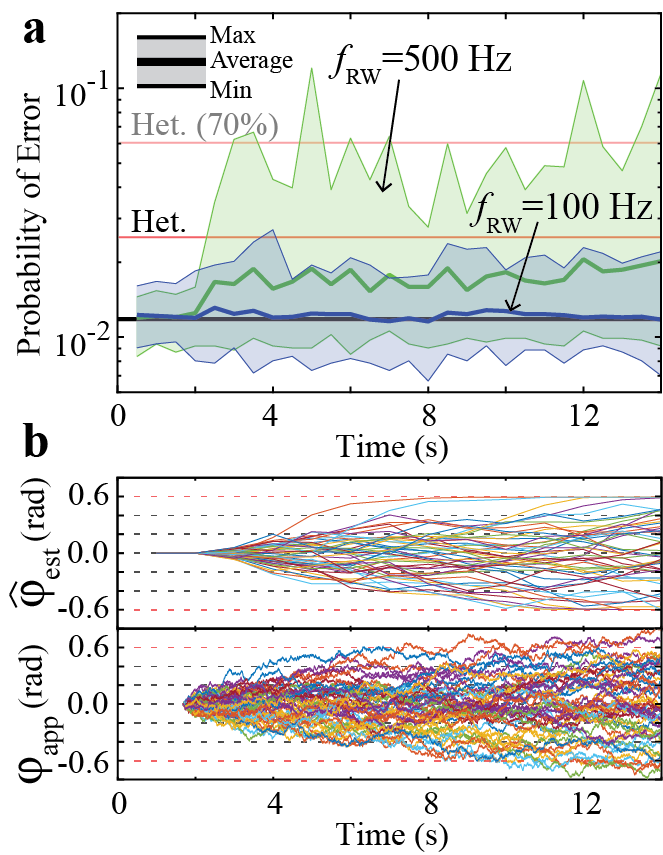}
\caption{\textbf{phase-tracking for different noise bandwidths.} (a) Probability of error $P_{\textrm{E}}$ as a function of time for $|\alpha|^{2} = 5.0$ for noise bandwidths $f_{\mathrm{RW}}$ = 100 Hz (blue) and $f_{\mathrm{RW}}$ = 500 Hz (green) with different random walks. Thick lines show the averages and shaded regions show the spread in $P_{\textrm{E}}$ over 50 Gaussian random walks. For $f_{\mathrm{RW}}$ = 100 Hz, the phase-tracking method allows the non-Gaussian receiver to perform 3.4 dB below the heterodyne limit (Het.), which is the expected performance in the absence of phase noise (black line). Noise with $f_{\mathrm{RW}}$ = 500 Hz causes $P_{\textrm{E}}$ to increase compared to $f_{\mathrm{RW}}$ = 100 Hz. However, phase-tracking allows receiver to perform below Het. for $~$ 14 s. (b) Applied random walks in phase for $f_{\mathrm{RW}}$ = 500 Hz (lower) and the phase-tracking estimates (upper). Applied phase $\phi_{\mathrm{app}}=\phi_{\mathrm{RW}}$ and estimates $\hat{\phi}_{\mathrm{est}}$ for $f_{\mathrm{RW}}$ = 100 Hz are shown in Fig. 4(e) in the main manuscript over 60 s.}
\label{mpn5fc}
\end{figure}

Figure \ref{mpn5fc}(a) shows the probability of error $P_{\textrm{E}}$ as a function of time from $t=0$ to 14s when applying random walks in phase with noise bandwidths $f_{\mathrm{RW}}$ = 100 Hz (blue) and $f_{\mathrm{RW}}$ = 500 Hz (green) for two different realizations of 50 random walks. Thick lines show the averages and shaded regions show the spread in $P_{\textrm{E}}$ over the 50 random walks. In this study, the random walks in phase and the phase-tracking are enabled at $t=2$ s in both cases [see Fig. \ref{mpn5fc}(b)] for the applied phase $\phi_{\mathrm{app}}$ and phase estimates $\hat{\phi}_{\mathrm{est}}$ for $f_{\mathrm{RW}} = 500$ Hz). We observe that in the presence of phase noise with $f_{\mathrm{RW}} = 100$ Hz, phase-tracking allows the receiver to maintain a 3.4 dB advantage over the ideal heterodyne limit (Het.) , which corresponds to the expected performance without phase noise \cite{becerra15} at $|\alpha|^{2} = 5.0$. On the other hand, for phase noise with bandwidth $f_{\mathrm{RW}}=500$ Hz, the average probability of error increases from this ideal case, showing a much larger spread in $P_{\textrm{E}}$ compared to the case with $f_{\mathrm{RW}}$ = 100 Hz. Performing phase-tracking for non-Gaussian receivers under phase noise with higher bandwidths $f_{\mathrm{RW}}$ requires achieving higher $f_{\mathrm{PT}}$ to generate accurate phase estimates $\hat{\phi}_{\mathrm{est}}$ at a sufficiently high rate compared to $f_{\mathrm{RW}}$. This can be achieved by increasing $f_{exp}$ resulting in higher sampling rates for phase estimation. Alternatively, increasing $|\alpha|^2$ would reduce errors in state discrimination and increase the accuracy of $\hat{\phi}_{\mathrm{est}}$, effectively reducing the effects of low phase-tracking bandwidths $f_{\mathrm{PT}}$. A combination of higher rates $f_{\mathrm{PT}}$ and low discrimination errors would make phase-tracking reliable for enabling the receiver to maintain discrimination below the heterodyne limit under phase noise with high noise bandwidths.
\section{Amplitude Noise}

In addition to phase noise, random amplitude fluctuations may occur in communication channels. Amplitude fluctuations will affect the discrimination strategy and will result in increased discrimination errors if the noise is relatively large. Table II shows the effects of amplitude fluctuations characterized by the standard deviation of the relative amplitude noise $\sigma_{amp}$ for $|\alpha|^{2}=5$ as a case study, which is assumed to be Gaussian such that $|\alpha|^{2} \rightarrow \mathcal{N}(1, \sigma_{amp})\times |\alpha|^{2}$. This increase in the probability of error for state discrimination causes higher errors in populating the probability distributions $P_{\delta}(n_{k}|\delta)$, from which the phase estimator $\hat{\phi}_{\mathrm{est}}$ is formed, and affects the phase-tracking performance, which can be characterized by the estimator variances. Table II shows the variances for the Sin-Cos and Bayesian estimators for different levels of amplitude noise. We observe that the presence of amplitude noise with levels from $\sigma_{amp}=5-25\%$ has very moderate effects on the error of state discrimination and on the variances of the estimators. This highlights the robustness of the phase-tracking method to amplitude noise.

\begin {table}
\begin{center}
\caption{Effect of amplitude noise in error discrimination and phase offset estimation for $|\alpha|^{2}=5$ and $N_{\mathrm{avg}} = 1$, including experimental imperfections.}
\begin{tabular}{ |c|c|c|c| }
 \hline
$\sigma_{amp}$ & $P_{\mathrm{E}} \times 10^{-2}$ & $\sigma_{Sin-Cos}^{2} \times 10^{-2}$ &  $\sigma_{Bayes}^{2} \times 10^{-2}$ \\
 \hline
 0.00 & 1.128 & 1.523 & 0.072 \\
 0.05 & 1.171 & 1.748 & 0.087 \\
 0.10 & 1.315 & 2.313 & 0.160 \\
 0.20 & 1.973 & 4.720 & 0.606 \\
 0.25 & 2.460 & 5.713 & 0.944 \\

 \hline
\end{tabular}
\end{center}
\end {table}

For situations with slow amplitude noise, it may be possible to perform amplitude estimation and tracking based on the collected data from state discrimination measurement. Specifically, the information contained in the photon number distributions $P_{\delta}(n_{k}|\delta)$ may be sufficient to estimate both phase and amplitude, which can eventually be used for phase and amplitude tracking. The receiver could use the Bayesian estimator for a two-dimensional estimation yielding a simultaneous estimate of phase offset and amplitude of the input state. Alternatively, the receiver could perform phase estimation with the method described here, and sequentially realize amplitude estimation based on Eq. (B1), or vice versa.


\begin{thebibliography}{70}
\expandafter\ifx\csname natexlab\endcsname\relax\def\natexlab#1{#1}\fi
\expandafter\ifx\csname bibnamefont\endcsname\relax
  \def\bibnamefont#1{#1}\fi
\expandafter\ifx\csname bibfnamefont\endcsname\relax
  \def\bibfnamefont#1{#1}\fi
\expandafter\ifx\csname citenamefont\endcsname\relax
  \def\citenamefont#1{#1}\fi
\expandafter\ifx\csname url\endcsname\relax
  \def\url#1{\texttt{#1}}\fi
\expandafter\ifx\csname urlprefix\endcsname\relax\def\urlprefix{URL }\fi
\providecommand{\bibinfo}[2]{#2}
\providecommand{\eprint}[2][]{\url{#2}}

\bibitem[{\citenamefont{Giovannetti et~al.}(2004)\citenamefont{Giovannetti,
  Guha, Lloyd, Maccone, Shapiro, and Yuen}}]{giovannetti04}
\bibinfo{author}{\bibfnamefont{V.}~\bibnamefont{Giovannetti}},
  \bibinfo{author}{\bibfnamefont{S.}~\bibnamefont{Guha}},
  \bibinfo{author}{\bibfnamefont{S.}~\bibnamefont{Lloyd}},
  \bibinfo{author}{\bibfnamefont{L.}~\bibnamefont{Maccone}},
  \bibinfo{author}{\bibfnamefont{J.~H.} \bibnamefont{Shapiro}},
  \bibnamefont{and} \bibinfo{author}{\bibfnamefont{H.~P.} \bibnamefont{Yuen}},
  \bibinfo{journal}{Phys. Rev. Lett.} \textbf{\bibinfo{volume}{92}},
  \bibinfo{pages}{027902} (\bibinfo{year}{2004}).

\bibitem[{\citenamefont{Giovannetti et~al.}(2013)\citenamefont{Giovannetti,
  Lloyd, and Maccone}}]{giovannetti11}
\bibinfo{author}{\bibfnamefont{V.}~\bibnamefont{Giovannetti}},
  \bibinfo{author}{\bibfnamefont{S.}~\bibnamefont{Lloyd}}, \bibnamefont{and}
  \bibinfo{author}{\bibfnamefont{L.}~\bibnamefont{Maccone}},
  \bibinfo{journal}{Nature Photon.} \textbf{\bibinfo{volume}{5}},
  \bibinfo{pages}{222} (\bibinfo{year}{2013}).

\bibitem[{\citenamefont{Mari et~al.}(2014)\citenamefont{Mari, Giovannetti, and
  Holevo}}]{mari14}
\bibinfo{author}{\bibfnamefont{A.}~\bibnamefont{Mari}},
  \bibinfo{author}{\bibfnamefont{V.}~\bibnamefont{Giovannetti}},
  \bibnamefont{and} \bibinfo{author}{\bibfnamefont{A.~S.}
  \bibnamefont{Holevo}}, \bibinfo{journal}{Nature Commun.}
  \textbf{\bibinfo{volume}{5}}, \bibinfo{pages}{3826} (\bibinfo{year}{2014}).

\bibitem[{\citenamefont{Winzer}(2012)}]{winzer12}
\bibinfo{author}{\bibfnamefont{P.~J.} \bibnamefont{Winzer}},
  \bibinfo{journal}{J. Lightwave Technol.} \textbf{\bibinfo{volume}{30}}
  (\bibinfo{year}{2012}).

\bibitem[{\citenamefont{Kikuchi}(2016)}]{kikuchi16}
\bibinfo{author}{\bibfnamefont{K.}~\bibnamefont{Kikuchi}},
  \bibinfo{journal}{Lightwave Technology, Journal of}
  \textbf{\bibinfo{volume}{34}}, \bibinfo{pages}{157} (\bibinfo{year}{2016}).

\bibitem[{\citenamefont{Ip et~al.}(2008)\citenamefont{Ip, Lau, Barros, and
  Kahn}}]{ip08}
\bibinfo{author}{\bibfnamefont{E.}~\bibnamefont{Ip}},
  \bibinfo{author}{\bibfnamefont{A.~P.~T.} \bibnamefont{Lau}},
  \bibinfo{author}{\bibfnamefont{D.~J.~F.} \bibnamefont{Barros}},
  \bibnamefont{and} \bibinfo{author}{\bibfnamefont{J.~M.} \bibnamefont{Kahn}},
  \bibinfo{journal}{Opt. Express} \textbf{\bibinfo{volume}{16}},
  \bibinfo{pages}{753} (\bibinfo{year}{2008}).

\bibitem[{\citenamefont{Li}(2009)}]{li09}
\bibinfo{author}{\bibfnamefont{G.}~\bibnamefont{Li}},
  \bibinfo{journal}{Advances in Optics and Photonics} pp.
  \textbf{\bibinfo{volume}{1}}, \bibinfo{pages}{279--307}
   (\bibinfo{year}{2009}).

\bibitem[{\citenamefont{Arrazola and L\"utkenhaus}(2014)}]{arrazola14}
\bibinfo{author}{\bibfnamefont{J.~M.} \bibnamefont{Arrazola}} \bibnamefont{and}
  \bibinfo{author}{\bibfnamefont{N.}~\bibnamefont{L\"utkenhaus}},
  \bibinfo{journal}{Phys. Rev. A} \textbf{\bibinfo{volume}{90}},
  \bibinfo{pages}{042335} (\bibinfo{year}{2014}).

\bibitem[{\citenamefont{Clarke et~al.}(2012)\citenamefont{Clarke, Collins,
  Dunjko, Andersson, Jeffers, and Buller}}]{clarke12}
\bibinfo{author}{\bibfnamefont{P.~J.} \bibnamefont{Clarke}},
  \bibinfo{author}{\bibfnamefont{R.~J.} \bibnamefont{Collins}},
  \bibinfo{author}{\bibfnamefont{V.}~\bibnamefont{Dunjko}},
  \bibinfo{author}{\bibfnamefont{E.}~\bibnamefont{Andersson}},
  \bibinfo{author}{\bibfnamefont{J.}~\bibnamefont{Jeffers}}, \bibnamefont{and}
  \bibinfo{author}{\bibfnamefont{G.~S.} \bibnamefont{Buller}},
  \bibinfo{journal}{Nat. Commun.} \textbf{\bibinfo{volume}{3}},
  \bibinfo{pages}{1174} (\bibinfo{year}{2012}).

\bibitem[{\citenamefont{Xu et~al.}(2015)\citenamefont{Xu, Arrazola, Wei, Wang,
  Palacios-Avila, Feng, Sajeed, Lutkenhaus, and Lo}}]{xu15}
\bibinfo{author}{\bibfnamefont{F.}~\bibnamefont{Xu}},
  \bibinfo{author}{\bibfnamefont{J.~M.} \bibnamefont{Arrazola}},
  \bibinfo{author}{\bibfnamefont{K.}~\bibnamefont{Wei}},
  \bibinfo{author}{\bibfnamefont{W.}~\bibnamefont{Wang}},
  \bibinfo{author}{\bibfnamefont{P.}~\bibnamefont{Palacios-Avila}},
  \bibinfo{author}{\bibfnamefont{C.}~\bibnamefont{Feng}},
  \bibinfo{author}{\bibfnamefont{S.}~\bibnamefont{Sajeed}},
  \bibinfo{author}{\bibfnamefont{N.}~\bibnamefont{Lutkenhaus}},
  \bibnamefont{and} \bibinfo{author}{\bibfnamefont{H.-K.} \bibnamefont{Lo}},
  \bibinfo{journal}{Nat. Commun.} \textbf{\bibinfo{volume}{6}},
  \bibinfo{pages}{8735} (\bibinfo{year}{2015}).

\bibitem[{\citenamefont{Bennett}(1992)}]{bennett92}
\bibinfo{author}{\bibfnamefont{C.~H.} \bibnamefont{Bennett}},
  \bibinfo{journal}{Phys. Rev. Lett.} \textbf{\bibinfo{volume}{68}},
  \bibinfo{pages}{3121} (\bibinfo{year}{1992}).

\bibitem[{\citenamefont{Huttner et~al.}(1995)\citenamefont{Huttner, Imoto,
  Gisin, and Mor}}]{huttner95}
\bibinfo{author}{\bibfnamefont{B.}~\bibnamefont{Huttner}},
  \bibinfo{author}{\bibfnamefont{N.}~\bibnamefont{Imoto}},
  \bibinfo{author}{\bibfnamefont{N.}~\bibnamefont{Gisin}}, \bibnamefont{and}
  \bibinfo{author}{\bibfnamefont{T.}~\bibnamefont{Mor}},
  \bibinfo{journal}{Phys. Rev. A} \textbf{\bibinfo{volume}{51}},
  \bibinfo{pages}{1863} (\bibinfo{year}{1995}).

\bibitem[{\citenamefont{Grosshans et~al.}(2003)\citenamefont{Grosshans, van
  Assche, Wenger, Tualle-Brouri, Cerf, and Grangier}}]{grosshans03}
\bibinfo{author}{\bibfnamefont{F.}~\bibnamefont{Grosshans}},
  \bibinfo{author}{\bibfnamefont{G.}~\bibnamefont{van Assche}},
  \bibinfo{author}{\bibfnamefont{J.}~\bibnamefont{Wenger}},
  \bibinfo{author}{\bibfnamefont{R.}~\bibnamefont{Tualle-Brouri}},
  \bibinfo{author}{\bibfnamefont{N.~J.} \bibnamefont{Cerf}}, \bibnamefont{and}
  \bibinfo{author}{\bibfnamefont{P.}~\bibnamefont{Grangier}},
  \bibinfo{journal}{Nature} \textbf{\bibinfo{volume}{421}},
  \bibinfo{pages}{238} (\bibinfo{year}{2003}).

\bibitem[{\citenamefont{Takeoka et~al.}(2014)\citenamefont{Takeoka, Guha, and
  Wilde}}]{takeoka14}
\bibinfo{author}{\bibfnamefont{M.}~\bibnamefont{Takeoka}},
  \bibinfo{author}{\bibfnamefont{S.}~\bibnamefont{Guha}}, \bibnamefont{and}
  \bibinfo{author}{\bibfnamefont{M.~M.} \bibnamefont{Wilde}},
  \bibinfo{journal}{Nat. Commun.} \textbf{\bibinfo{volume}{5}},
  \bibinfo{pages}{5235} (\bibinfo{year}{2014}).

\bibitem[{\citenamefont{Pirandola et~al.}(2017)\citenamefont{Pirandola,
  Laurenza, Ottaviani, and Banchi}}]{pirandola17}
\bibinfo{author}{\bibfnamefont{S.}~\bibnamefont{Pirandola}},
  \bibinfo{author}{\bibfnamefont{R.}~\bibnamefont{Laurenza}},
  \bibinfo{author}{\bibfnamefont{C.}~\bibnamefont{Ottaviani}},
  \bibnamefont{and} \bibinfo{author}{\bibfnamefont{L.}~\bibnamefont{Banchi}},
  \bibinfo{journal}{Nat. Commun.} \textbf{\bibinfo{volume}{8}},
  \bibinfo{pages}{15043} (\bibinfo{year}{2017}).

\bibitem[{\citenamefont{Ghorai et~al.}(2019)\citenamefont{Ghorai, Grangier,
  Diamanti, and Leverrier}}]{ghorai19}
\bibinfo{author}{\bibfnamefont{S.}~\bibnamefont{Ghorai}},
  \bibinfo{author}{\bibfnamefont{P.}~\bibnamefont{Grangier}},
  \bibinfo{author}{\bibfnamefont{E.}~\bibnamefont{Diamanti}}, \bibnamefont{and}
  \bibinfo{author}{\bibfnamefont{A.}~\bibnamefont{Leverrier}},
  \bibinfo{journal}{Phys. Rev. X} \textbf{\bibinfo{volume}{9}},
  \bibinfo{pages}{021059} (\bibinfo{year}{2019}).

\bibitem[{\citenamefont{Tang et~al.}(2016)\citenamefont{Tang, Yin, Zhao, Liu,
  Sun, Huang, Zhang, Chen, Zhang, You et~al.}}]{tang16}
\bibinfo{author}{\bibfnamefont{Y.-L.} \bibnamefont{Tang}},
  \bibinfo{author}{\bibfnamefont{H.-L.} \bibnamefont{Yin}},
  \bibinfo{author}{\bibfnamefont{Q.}~\bibnamefont{Zhao}},
  \bibinfo{author}{\bibfnamefont{H.}~\bibnamefont{Liu}},
  \bibinfo{author}{\bibfnamefont{X.-X.} \bibnamefont{Sun}},
  \bibinfo{author}{\bibfnamefont{M.-Q.} \bibnamefont{Huang}},
  \bibinfo{author}{\bibfnamefont{W.-J.} \bibnamefont{Zhang}},
  \bibinfo{author}{\bibfnamefont{S.-J.} \bibnamefont{Chen}},
  \bibinfo{author}{\bibfnamefont{L.}~\bibnamefont{Zhang}},
  \bibinfo{author}{\bibfnamefont{L.-X.} \bibnamefont{You}},
  \bibnamefont{et~al.}, \bibinfo{journal}{Phys. Rev. X}
  \textbf{\bibinfo{volume}{6}}, \bibinfo{pages}{011024} (\bibinfo{year}{2016}).

\bibitem[{\citenamefont{Sasaki et~al.}(2011)\citenamefont{Sasaki, Fujiwara,
  Ishizuka, Klaus, Wakui, Takeoka, Miki, Yamashita, Wang, Tanaka
  et~al.}}]{sasaki11}
\bibinfo{author}{\bibfnamefont{M.}~\bibnamefont{Sasaki}},
  \bibinfo{author}{\bibfnamefont{M.}~\bibnamefont{Fujiwara}},
  \bibinfo{author}{\bibfnamefont{H.}~\bibnamefont{Ishizuka}},
  \bibinfo{author}{\bibfnamefont{W.}~\bibnamefont{Klaus}},
  \bibinfo{author}{\bibfnamefont{K.}~\bibnamefont{Wakui}},
  \bibinfo{author}{\bibfnamefont{M.}~\bibnamefont{Takeoka}},
  \bibinfo{author}{\bibfnamefont{S.}~\bibnamefont{Miki}},
  \bibinfo{author}{\bibfnamefont{T.}~\bibnamefont{Yamashita}},
  \bibinfo{author}{\bibfnamefont{Z.}~\bibnamefont{Wang}},
  \bibinfo{author}{\bibfnamefont{A.}~\bibnamefont{Tanaka}},
  \bibnamefont{et~al.}, \bibinfo{journal}{Opt. Express}
  \textbf{\bibinfo{volume}{19}} pp.
  \bibinfo{pages}{10387--10409} (\bibinfo{year}{2011}).

\bibitem[{\citenamefont{Armada and Calvo}(1998)}]{armada98}
\bibinfo{author}{\bibfnamefont{A.~G.} \bibnamefont{Armada}} \bibnamefont{and}
  \bibinfo{author}{\bibfnamefont{M.}~\bibnamefont{Calvo}},
  \bibinfo{journal}{IEEE Communications Letters} \textbf{\bibinfo{volume}{2}},
  \textbf{\bibinfo{pages}{11}}
  (\bibinfo{year}{1998}).

\bibitem[{\citenamefont{Qi et~al.}(2007)\citenamefont{Qi, Huang, Qian, and
  Lo}}]{qi07}
\bibinfo{author}{\bibfnamefont{B.}~\bibnamefont{Qi}},
  \bibinfo{author}{\bibfnamefont{L.-L.} \bibnamefont{Huang}},
  \bibinfo{author}{\bibfnamefont{L.}~\bibnamefont{Qian}}, \bibnamefont{and}
  \bibinfo{author}{\bibfnamefont{H.-K.} \bibnamefont{Lo}},
  \bibinfo{journal}{Phys. Rev. A} \textbf{\bibinfo{volume}{76}},
  \bibinfo{pages}{052323} (\bibinfo{year}{2007}).

\bibitem[{\citenamefont{Jouguet et~al.}(2103)\citenamefont{Jouguet,
  Kunz-Jacques, Leverrier, Grangier, and Diamanti}}]{jouget13}
\bibinfo{author}{\bibfnamefont{P.}~\bibnamefont{Jouguet}},
  \bibinfo{author}{\bibfnamefont{S.}~\bibnamefont{Kunz-Jacques}},
  \bibinfo{author}{\bibfnamefont{A.}~\bibnamefont{Leverrier}},
  \bibinfo{author}{\bibfnamefont{P.}~\bibnamefont{Grangier}}, \bibnamefont{and}
  \bibinfo{author}{\bibfnamefont{E.}~\bibnamefont{Diamanti}},
  \bibinfo{journal}{Nature Photonics} \textbf{\bibinfo{volume}{7}},
  \bibinfo{pages}{378} (\bibinfo{year}{2103}).

\bibitem[{\citenamefont{Qi et~al.}(2015)\citenamefont{Qi, Lougovski, Pooser,
  Grice, and Bobrek}}]{qi15}
\bibinfo{author}{\bibfnamefont{B.}~\bibnamefont{Qi}},
  \bibinfo{author}{\bibfnamefont{P.}~\bibnamefont{Lougovski}},
  \bibinfo{author}{\bibfnamefont{R.}~\bibnamefont{Pooser}},
  \bibinfo{author}{\bibfnamefont{W.}~\bibnamefont{Grice}}, \bibnamefont{and}
  \bibinfo{author}{\bibfnamefont{M.}~\bibnamefont{Bobrek}},
  \bibinfo{journal}{Phys. Rev. X} \textbf{\bibinfo{volume}{5}},
  \bibinfo{pages}{041009} (\bibinfo{year}{2015}).

\bibitem[{\citenamefont{Soh et~al.}(2015)\citenamefont{Soh, Brif, Coles,
  L\"utkenhaus, Camacho, Urayama, and Sarovar}}]{soh15}
\bibinfo{author}{\bibfnamefont{D.~B.~S.} \bibnamefont{Soh}},
  \bibinfo{author}{\bibfnamefont{C.}~\bibnamefont{Brif}},
  \bibinfo{author}{\bibfnamefont{P.~J.} \bibnamefont{Coles}},
  \bibinfo{author}{\bibfnamefont{N.}~\bibnamefont{L\"utkenhaus}},
  \bibinfo{author}{\bibfnamefont{R.~M.} \bibnamefont{Camacho}},
  \bibinfo{author}{\bibfnamefont{J.}~\bibnamefont{Urayama}}, \bibnamefont{and}
  \bibinfo{author}{\bibfnamefont{M.}~\bibnamefont{Sarovar}},
  \bibinfo{journal}{Phys. Rev. X} \textbf{\bibinfo{volume}{5}},
  \bibinfo{pages}{041010} (\bibinfo{year}{2015}).

\bibitem[{\citenamefont{Huang et~al.}(2015)\citenamefont{Huang, Huang, Lin,
  Wang, and Zeng}}]{huang15}
\bibinfo{author}{\bibfnamefont{D.}~\bibnamefont{Huang}},
  \bibinfo{author}{\bibfnamefont{P.}~\bibnamefont{Huang}},
  \bibinfo{author}{\bibfnamefont{D.}~\bibnamefont{Lin}},
  \bibinfo{author}{\bibfnamefont{C.}~\bibnamefont{Wang}}, \bibnamefont{and}
  \bibinfo{author}{\bibfnamefont{G.}~\bibnamefont{Zeng}},
  \bibinfo{journal}{Opt. Lett.} \textbf{\bibinfo{volume}{40}},
  \bibinfo{pages}{3695} (\bibinfo{year}{2015}).

\bibitem[{\citenamefont{Marie and All\'eaume}(2017)}]{marie17}
\bibinfo{author}{\bibfnamefont{A.}~\bibnamefont{Marie}} \bibnamefont{and}
  \bibinfo{author}{\bibfnamefont{R.}~\bibnamefont{All\'eaume}},
  \bibinfo{journal}{Phys. Rev. A} \textbf{\bibinfo{volume}{95}},
  \bibinfo{pages}{012316} (\bibinfo{year}{2017}).

\bibitem[{\citenamefont{Barry and Kahn}(1992)}]{barry92}
\bibinfo{author}{\bibfnamefont{J.~R.} \bibnamefont{Barry}} \bibnamefont{and}
  \bibinfo{author}{\bibfnamefont{J.~M.} \bibnamefont{Kahn}},
  \bibinfo{journal}{Journal of Lightwave Technology}
  \textbf{\bibinfo{volume}{10}} (\bibinfo{year}{1992}).

\bibitem[{\citenamefont{Ly-Gagnon et~al.}(2006)\citenamefont{Ly-Gagnon,
  Tsukamoto, Katoh, and Kikuchi}}]{lygagnon06}
\bibinfo{author}{\bibfnamefont{D.~S.} \bibnamefont{Ly-Gagnon}},
  \bibinfo{author}{\bibfnamefont{S.}~\bibnamefont{Tsukamoto}},
  \bibinfo{author}{\bibfnamefont{K.}~\bibnamefont{Katoh}}, \bibnamefont{and}
  \bibinfo{author}{\bibfnamefont{K.}~\bibnamefont{Kikuchi}},
  \bibinfo{journal}{Journal of Lightwave Technology}
  \textbf{\bibinfo{volume}{24}} (\bibinfo{year}{2006}).

\bibitem[{\citenamefont{Ip and Kahn}(2007)}]{ip07}
\bibinfo{author}{\bibfnamefont{E.}~\bibnamefont{Ip}} \bibnamefont{and}
  \bibinfo{author}{\bibfnamefont{J.~M.} \bibnamefont{Kahn}},
  \bibinfo{journal}{Journal of Lightwave Technology}
  \textbf{\bibinfo{volume}{25}}, \textbf{\bibinfo{pages}{2675}}
   (\bibinfo{year}{2007}).

\bibitem[{\citenamefont{Morsy-Osman et~al.}(2011)\citenamefont{Morsy-Osman,
  Zhuge, Chen, and Plant}}]{morsyosman11}
\bibinfo{author}{\bibfnamefont{M.}~\bibnamefont{Morsy-Osman}},
  \bibinfo{author}{\bibfnamefont{Q.}~\bibnamefont{Zhuge}},
  \bibinfo{author}{\bibfnamefont{L.~R.} \bibnamefont{Chen}}, \bibnamefont{and}
  \bibinfo{author}{\bibfnamefont{D.~V.} \bibnamefont{Plant}},
  \bibinfo{journal}{Opt. Express} \textbf{\bibinfo{volume}{19}},
  \bibinfo{pages}{B329} (\bibinfo{year}{2011}).

\bibitem[{\citenamefont{Wang et~al.}(2019)\citenamefont{Wang, Huang, Wang, and
  Zeng}}]{wang19}
\bibinfo{author}{\bibfnamefont{T.}~\bibnamefont{Wang}},
  \bibinfo{author}{\bibfnamefont{P.}~\bibnamefont{Huang}},
  \bibinfo{author}{\bibfnamefont{S.}~\bibnamefont{Wang}}, \bibnamefont{and}
  \bibinfo{author}{\bibfnamefont{G.}~\bibnamefont{Zeng}},
  \bibinfo{journal}{Phys. Rev. A} \textbf{\bibinfo{volume}{99}},
  \bibinfo{pages}{022318} (\bibinfo{year}{2019}).

\bibitem[{\citenamefont{He et~al.}(2014)\citenamefont{He, Norwood,
  Brandt-Pearce, Djordjevic, Cvijetic, Subramaniam, Himmelhuber, Reynolds,
  Blanche, Lynn et~al.}}]{he14}
\bibinfo{author}{\bibfnamefont{J.}~\bibnamefont{He}},
  \bibinfo{author}{\bibfnamefont{R.}~\bibnamefont{Norwood}},
  \bibinfo{author}{\bibfnamefont{M.}~\bibnamefont{Brandt-Pearce}},
  \bibinfo{author}{\bibfnamefont{I.}~\bibnamefont{Djordjevic}},
  \bibinfo{author}{\bibfnamefont{M.}~\bibnamefont{Cvijetic}},
  \bibinfo{author}{\bibfnamefont{S.}~\bibnamefont{Subramaniam}},
  \bibinfo{author}{\bibfnamefont{R.}~\bibnamefont{Himmelhuber}},
  \bibinfo{author}{\bibfnamefont{C.}~\bibnamefont{Reynolds}},
  \bibinfo{author}{\bibfnamefont{P.}~\bibnamefont{Blanche}},
  \bibinfo{author}{\bibfnamefont{B.}~\bibnamefont{Lynn}}, \bibnamefont{et~al.},
  \bibinfo{journal}{Comp. and Elec. Eng.} \textbf{\bibinfo{volume}{40}},
  \bibinfo{pages}{216} (\bibinfo{year}{2014}).

\bibitem[{\citenamefont{Agrell et~al.}(2016)\citenamefont{Agrell, Karlsson,
  Chraplyvy, Richardson, Krummrich, Winzer, Roberts, Fischer, Savory, Eggleton
  et~al.}}]{gisin16}
\bibinfo{author}{\bibfnamefont{E.}~\bibnamefont{Agrell}},
  \bibinfo{author}{\bibfnamefont{M.}~\bibnamefont{Karlsson}},
  \bibinfo{author}{\bibfnamefont{A.~R.} \bibnamefont{Chraplyvy}},
  \bibinfo{author}{\bibfnamefont{D.~J.} \bibnamefont{Richardson}},
  \bibinfo{author}{\bibfnamefont{P.~M.} \bibnamefont{Krummrich}},
  \bibinfo{author}{\bibfnamefont{P.}~\bibnamefont{Winzer}},
  \bibinfo{author}{\bibfnamefont{K.}~\bibnamefont{Roberts}},
  \bibinfo{author}{\bibfnamefont{J.~K.} \bibnamefont{Fischer}},
  \bibinfo{author}{\bibfnamefont{S.~J.} \bibnamefont{Savory}},
  \bibinfo{author}{\bibfnamefont{B.~J.} \bibnamefont{Eggleton}},
  \bibnamefont{et~al.}, \bibinfo{journal}{Jour. of Opt.}
  \textbf{\bibinfo{volume}{18}}, \bibinfo{pages}{063002}
  (\bibinfo{year}{2016}).

\bibitem[{\citenamefont{Helstrom}(1976)}]{helstrom76}
\bibinfo{author}{\bibfnamefont{C.~W.} \bibnamefont{Helstrom}},
  \emph{\bibinfo{title}{Quantum detection and estimation theory, Mathematics in
  Science and Engineering Vol. 123}} (\bibinfo{publisher}{Academic Press},
  \bibinfo{address}{New York}, \bibinfo{year}{1976}).

\bibitem[{\citenamefont{Weedbrook et~al.}(2012)\citenamefont{Weedbrook,
  Pirandola, Garc\'ia-Patr\'on, Cerf, Ralph, Shapiro, and Lloyd}}]{weedbrook12}
\bibinfo{author}{\bibfnamefont{C.}~\bibnamefont{Weedbrook}},
  \bibinfo{author}{\bibfnamefont{S.}~\bibnamefont{Pirandola}},
  \bibinfo{author}{\bibfnamefont{R.}~\bibnamefont{Garc\'ia-Patr\'on}},
  \bibinfo{author}{\bibfnamefont{N.~J.} \bibnamefont{Cerf}},
  \bibinfo{author}{\bibfnamefont{T.~C.} \bibnamefont{Ralph}},
  \bibinfo{author}{\bibfnamefont{J.~H.} \bibnamefont{Shapiro}},
  \bibnamefont{and} \bibinfo{author}{\bibfnamefont{S.}~\bibnamefont{Lloyd}},
  \bibinfo{journal}{Rev. Mod. Phys.} \textbf{\bibinfo{volume}{84}},
  \bibinfo{pages}{621} (\bibinfo{year}{2012}).

\bibitem[{\citenamefont{Wittmann
  et~al.}(2010{\natexlab{a}})\citenamefont{Wittmann, Andersen, Takeoka, Sych,
  and Leuchs}}]{wittmann10}
\bibinfo{author}{\bibfnamefont{C.}~\bibnamefont{Wittmann}},
  \bibinfo{author}{\bibfnamefont{U.~L.} \bibnamefont{Andersen}},
  \bibinfo{author}{\bibfnamefont{M.}~\bibnamefont{Takeoka}},
  \bibinfo{author}{\bibfnamefont{D.}~\bibnamefont{Sych}}, \bibnamefont{and}
  \bibinfo{author}{\bibfnamefont{G.}~\bibnamefont{Leuchs}},
  \bibinfo{journal}{Phys. Rev. Lett.} \textbf{\bibinfo{volume}{104}},
  \bibinfo{pages}{100505} (\bibinfo{year}{2010}{\natexlab{a}}).

\bibitem[{\citenamefont{Wiseman}(1995)}]{wiseman95}
\bibinfo{author}{\bibfnamefont{H.~M.} \bibnamefont{Wiseman}},
  \bibinfo{journal}{Phys. Rev. Lett.} \textbf{\bibinfo{volume}{75}},
  \bibinfo{pages}{4587} (\bibinfo{year}{1995}).

\bibitem[{\citenamefont{Wiseman and Killip}(1998)}]{wiseman98}
\bibinfo{author}{\bibfnamefont{H.~M.} \bibnamefont{Wiseman}} \bibnamefont{and}
  \bibinfo{author}{\bibfnamefont{R.~B.} \bibnamefont{Killip}},
  \bibinfo{journal}{Phys. Rev. A} \textbf{\bibinfo{volume}{57}},
  \bibinfo{pages}{2169} (\bibinfo{year}{1998}).

\bibitem[{\citenamefont{D'Ariano et~al.}(1996)\citenamefont{D'Ariano, Paris,
  and Seno}}]{giacomo96}
\bibinfo{author}{\bibfnamefont{G.~M.} \bibnamefont{D'Ariano}},
  \bibinfo{author}{\bibfnamefont{M.~G.~A.} \bibnamefont{Paris}},
  \bibnamefont{and} \bibinfo{author}{\bibfnamefont{R.}~\bibnamefont{Seno}},
  \bibinfo{journal}{Phys. Rev. A} \textbf{\bibinfo{volume}{54}},
  \bibinfo{pages}{4495} (\bibinfo{year}{1996}).

\bibitem[{\citenamefont{Chesi et~al.}(2018)\citenamefont{Chesi, Olivares, and
  Paris}}]{chesi18}
\bibinfo{author}{\bibfnamefont{G.}~\bibnamefont{Chesi}},
  \bibinfo{author}{\bibfnamefont{S.}~\bibnamefont{Olivares}}, \bibnamefont{and}
  \bibinfo{author}{\bibfnamefont{M.~G.~A.} \bibnamefont{Paris}},
  \bibinfo{journal}{Phys. Rev. A} \textbf{\bibinfo{volume}{97}},
  \bibinfo{pages}{032315} (\bibinfo{year}{2018}).

\bibitem[{\citenamefont{Kennedy}()}]{kennedy72}
\bibinfo{author}{\bibfnamefont{R.~S.} \bibnamefont{Kennedy}},
  \emph{\bibinfo{title}{A Near-Optimum Receiver for the Binary Coherent State Quantum Channel.}}, \bibinfo{note}{MIT Research Laboratory of Electronics Quarterly Progress Report 108: 219-225 (1973), unpublished}.

\bibitem[{\citenamefont{Dolinar}()}]{dolinar73}
\bibinfo{author}{\bibfnamefont{S.~J.} \bibnamefont{Dolinar}},
  \emph{\bibinfo{title}{An optimum receiver for the binary coherent state
  quantum channel}}, \bibinfo{note}{research Laboratory of Electronics, MIT,
  Quarterly Progress Report No. 111 (1973), p. 115.}

\bibitem[{\citenamefont{Bondurant}(1993)}]{bondurant93}
\bibinfo{author}{\bibfnamefont{R.~S.} \bibnamefont{Bondurant}},
  \bibinfo{journal}{Opt. Lett.} \textbf{\bibinfo{volume}{18}},
  \bibinfo{pages}{1896} (\bibinfo{year}{1993}).

\bibitem[{\citenamefont{Cook et~al.}(2007)\citenamefont{Cook, Martin, and
  Geremia}}]{cook07}
\bibinfo{author}{\bibfnamefont{R.~L.} \bibnamefont{Cook}},
  \bibinfo{author}{\bibfnamefont{P.~J.} \bibnamefont{Martin}},
  \bibnamefont{and} \bibinfo{author}{\bibfnamefont{J.~M.}
  \bibnamefont{Geremia}}, \bibinfo{journal}{Nature}
  \textbf{\bibinfo{volume}{446}}, \bibinfo{pages}{774} (\bibinfo{year}{2007}).

\bibitem[{\citenamefont{Wittmann et~al.}(2008)\citenamefont{Wittmann, Takeoka,
  Cassemiro, Sasaki, Leuchs, and Andersen}}]{wittman08}
\bibinfo{author}{\bibfnamefont{C.}~\bibnamefont{Wittmann}},
  \bibinfo{author}{\bibfnamefont{M.}~\bibnamefont{Takeoka}},
  \bibinfo{author}{\bibfnamefont{K.~N.} \bibnamefont{Cassemiro}},
  \bibinfo{author}{\bibfnamefont{M.}~\bibnamefont{Sasaki}},
  \bibinfo{author}{\bibfnamefont{G.}~\bibnamefont{Leuchs}}, \bibnamefont{and}
  \bibinfo{author}{\bibfnamefont{U.~L.} \bibnamefont{Andersen}},
  \bibinfo{journal}{Phys. Rev. Lett.} \textbf{\bibinfo{volume}{101}},
  \bibinfo{pages}{210501} (\bibinfo{year}{2008}).

\bibitem[{\citenamefont{Wittmann
  et~al.}(2010{\natexlab{b}})\citenamefont{Wittmann, Andersen, and
  Leuchs}}]{wittmann10b}
\bibinfo{author}{\bibfnamefont{C.}~\bibnamefont{Wittmann}},
  \bibinfo{author}{\bibfnamefont{U.~L.} \bibnamefont{Andersen}},
  \bibnamefont{and} \bibinfo{author}{\bibfnamefont{G.}~\bibnamefont{Leuchs}},
  \bibinfo{journal}{Journal of Modern Optics} \textbf{\bibinfo{volume}{57}},
  \bibinfo{pages}{213} (\bibinfo{year}{2010}{\natexlab{b}}).

\bibitem[{\citenamefont{Tsujino et~al.}(2011)\citenamefont{Tsujino, Fukuda,
  Fujii, Inoue, Fujiwara, Takeoka, and Sasaki}}]{tsujino11}
\bibinfo{author}{\bibfnamefont{K.}~\bibnamefont{Tsujino}},
  \bibinfo{author}{\bibfnamefont{D.}~\bibnamefont{Fukuda}},
  \bibinfo{author}{\bibfnamefont{G.}~\bibnamefont{Fujii}},
  \bibinfo{author}{\bibfnamefont{S.}~\bibnamefont{Inoue}},
  \bibinfo{author}{\bibfnamefont{M.}~\bibnamefont{Fujiwara}},
  \bibinfo{author}{\bibfnamefont{M.}~\bibnamefont{Takeoka}}, \bibnamefont{and}
  \bibinfo{author}{\bibfnamefont{M.}~\bibnamefont{Sasaki}},
  \bibinfo{journal}{Phys. Rev. Lett.} \textbf{\bibinfo{volume}{106}},
  \bibinfo{pages}{250503} (\bibinfo{year}{2011}).

\bibitem[{\citenamefont{Becerra et~al.}(2011)\citenamefont{Becerra, Fan,
  Baumgartner, Polyakov, Goldhar, Kosloski, and Migdall}}]{becerra11}
\bibinfo{author}{\bibfnamefont{F.~E.} \bibnamefont{Becerra}},
  \bibinfo{author}{\bibfnamefont{J.}~\bibnamefont{Fan}},
  \bibinfo{author}{\bibfnamefont{G.}~\bibnamefont{Baumgartner}},
  \bibinfo{author}{\bibfnamefont{S.~V.} \bibnamefont{Polyakov}},
  \bibinfo{author}{\bibfnamefont{J.}~\bibnamefont{Goldhar}},
  \bibinfo{author}{\bibfnamefont{J.~T.} \bibnamefont{Kosloski}},
  \bibnamefont{and} \bibinfo{author}{\bibfnamefont{A.}~\bibnamefont{Migdall}},
  \bibinfo{journal}{Phys. Rev. A} \textbf{\bibinfo{volume}{84}},
  \bibinfo{pages}{062324} (\bibinfo{year}{2011}).

\bibitem[{\citenamefont{M\"{u}ller et~al.}(2012)\citenamefont{M\"{u}ller,
  Usuga, Wittmann, Takeoka, Marquardt, Andersen, and Leuchs}}]{muller12}
\bibinfo{author}{\bibfnamefont{C.~R.} \bibnamefont{M\"{u}ller}},
  \bibinfo{author}{\bibfnamefont{M.~A.} \bibnamefont{Usuga}},
  \bibinfo{author}{\bibfnamefont{C.}~\bibnamefont{Wittmann}},
  \bibinfo{author}{\bibfnamefont{M.}~\bibnamefont{Takeoka}},
  \bibinfo{author}{\bibfnamefont{C.}~\bibnamefont{Marquardt}},
  \bibinfo{author}{\bibfnamefont{U.~L.} \bibnamefont{Andersen}},
  \bibnamefont{and} \bibinfo{author}{\bibfnamefont{G.}~\bibnamefont{Leuchs}},
  \bibinfo{journal}{New J. of Phys.} \textbf{\bibinfo{volume}{14}},
  \bibinfo{pages}{083009} (\bibinfo{year}{2012}).

\bibitem[{\citenamefont{Becerra et~al.}(2013)\citenamefont{Becerra, Fan,
  Baumgartner, Goldhar, Kosloski, and Migdall}}]{becerra13}
\bibinfo{author}{\bibfnamefont{F.~E.} \bibnamefont{Becerra}},
  \bibinfo{author}{\bibfnamefont{J.}~\bibnamefont{Fan}},
  \bibinfo{author}{\bibfnamefont{G.}~\bibnamefont{Baumgartner}},
  \bibinfo{author}{\bibfnamefont{J.}~\bibnamefont{Goldhar}},
  \bibinfo{author}{\bibfnamefont{J.~T.} \bibnamefont{Kosloski}},
  \bibnamefont{and} \bibinfo{author}{\bibfnamefont{A.}~\bibnamefont{Migdall}},
  \bibinfo{journal}{Nature Photonics} \textbf{\bibinfo{volume}{7}},
  \bibinfo{pages}{147} (\bibinfo{year}{2013}).

\bibitem[{\citenamefont{Izumi et~al.}(2013)\citenamefont{Izumi, Takeoka, Ema,
  and Sasaki}}]{izumi13}
\bibinfo{author}{\bibfnamefont{S.}~\bibnamefont{Izumi}},
  \bibinfo{author}{\bibfnamefont{M.}~\bibnamefont{Takeoka}},
  \bibinfo{author}{\bibfnamefont{K.}~\bibnamefont{Ema}}, \bibnamefont{and}
  \bibinfo{author}{\bibfnamefont{M.}~\bibnamefont{Sasaki}},
  \bibinfo{journal}{Phys. Rev. A} \textbf{\bibinfo{volume}{87}},
  \bibinfo{pages}{042328} (\bibinfo{year}{2013}).

\bibitem[{\citenamefont{Nair et~al.}(2014)\citenamefont{Nair, Guha, and
  Tan}}]{nair14}
\bibinfo{author}{\bibfnamefont{R.}~\bibnamefont{Nair}},
  \bibinfo{author}{\bibfnamefont{S.}~\bibnamefont{Guha}}, \bibnamefont{and}
  \bibinfo{author}{\bibfnamefont{S.-H.} \bibnamefont{Tan}},
  \bibinfo{journal}{Phys. Rev. A} \textbf{\bibinfo{volume}{89}},
  \bibinfo{pages}{032318} (\bibinfo{year}{2014}).

\bibitem[{\citenamefont{M\"{u}ller and Marquardt}(2015)}]{muller15}
\bibinfo{author}{\bibfnamefont{C.~R.} \bibnamefont{M\"{u}ller}}
  \bibnamefont{and}
  \bibinfo{author}{\bibfnamefont{C.}~\bibnamefont{Marquardt}},
  \bibinfo{journal}{New Journal of Physics} \textbf{\bibinfo{volume}{17}},
  \bibinfo{pages}{032003} (\bibinfo{year}{2015}).

\bibitem[{\citenamefont{Becerra et~al.}(2015)\citenamefont{Becerra, Fan, and
  Migdall}}]{becerra15}
\bibinfo{author}{\bibfnamefont{F.~E.} \bibnamefont{Becerra}},
  \bibinfo{author}{\bibfnamefont{J.}~\bibnamefont{Fan}}, \bibnamefont{and}
  \bibinfo{author}{\bibfnamefont{A.}~\bibnamefont{Migdall}},
  \bibinfo{journal}{Nat. Photonics} \textbf{\bibinfo{volume}{9}}
  (\bibinfo{year}{2015}).

\bibitem[{\citenamefont{Bina et~al.}(2016)\citenamefont{Bina, Allevi, Bondani,
  and Olivares}}]{bina16}
\bibinfo{author}{\bibfnamefont{M.}~\bibnamefont{Bina}},
  \bibinfo{author}{\bibfnamefont{A.}~\bibnamefont{Allevi}},
  \bibinfo{author}{\bibfnamefont{M.}~\bibnamefont{Bondani}}, \bibnamefont{and}
  \bibinfo{author}{\bibfnamefont{S.}~\bibnamefont{Olivares}},
  \bibinfo{journal}{Scientific Reports}, \textbf{\bibinfo{volume}{6}}
  \bibinfo{pages}{26025}
  (\bibinfo{year}{2016}).

\bibitem[{\citenamefont{Ferdinand et~al.}(2017)\citenamefont{Ferdinand,
  DiMario, and Becerra}}]{ferdinand17}
\bibinfo{author}{\bibfnamefont{A.~R.} \bibnamefont{Ferdinand}},
  \bibinfo{author}{\bibfnamefont{M.~T.} \bibnamefont{DiMario}},
  \bibnamefont{and} \bibinfo{author}{\bibfnamefont{F.~E.}
  \bibnamefont{Becerra}}, \bibinfo{journal}{npj Quantum Information}
  \textbf{\bibinfo{volume}{3}}, \bibinfo{pages}{43} (\bibinfo{year}{2017}).

\bibitem[{\citenamefont{DiMario et~al.}(2018)\citenamefont{DiMario, Carrasco,
  Jackson, and Becerra}}]{dimario18}
\bibinfo{author}{\bibfnamefont{M.~T.} \bibnamefont{DiMario}},
  \bibinfo{author}{\bibfnamefont{E.}~\bibnamefont{Carrasco}},
  \bibinfo{author}{\bibfnamefont{R.~A.} \bibnamefont{Jackson}},
  \bibnamefont{and} \bibinfo{author}{\bibfnamefont{F.~E.}
  \bibnamefont{Becerra}}, \bibinfo{journal}{J. Opt. Soc. Am. B}
  \textbf{\bibinfo{volume}{35}}, \bibinfo{pages}{568} (\bibinfo{year}{2018}).

\bibitem[{\citenamefont{DiMario and Becerra}(2018)}]{dimario18b}
\bibinfo{author}{\bibfnamefont{M.~T.} \bibnamefont{DiMario}} \bibnamefont{and}
  \bibinfo{author}{\bibfnamefont{F.~E.} \bibnamefont{Becerra}},
  \bibinfo{journal}{Phys. Rev. Lett.} \textbf{\bibinfo{volume}{121}},
  \bibinfo{pages}{023603} (\bibinfo{year}{2018}).

\bibitem[{\citenamefont{DiMario et~al.}(0000)\citenamefont{DiMario, Kunz,
  Banaszek, and Becerra}}]{dimario19}
\bibinfo{author}{\bibfnamefont{M.~T.} \bibnamefont{DiMario}},
  \bibinfo{author}{\bibfnamefont{L.}~\bibnamefont{Kunz}},
  \bibinfo{author}{\bibfnamefont{K.}~\bibnamefont{Banaszek}}, \bibnamefont{and}
  \bibinfo{author}{\bibfnamefont{F.~E.} \bibnamefont{Becerra}},
  \bibinfo{journal}{npj Quantum Information} \textbf{\bibinfo{volume}{5}},
  \bibinfo{pages}{65} (\bibinfo{year}{2019}).

\bibitem[{\citenamefont{Guha}(2011)}]{guha11}
\bibinfo{author}{\bibfnamefont{S.}~\bibnamefont{Guha}}, \bibinfo{journal}{Phys.
  Rev. Lett.} \textbf{\bibinfo{volume}{106}}, \bibinfo{pages}{240502}
  (\bibinfo{year}{2011}).

\bibitem[{\citenamefont{Izumi et~al.}(2016)\citenamefont{Izumi, Takeoka, Wakui,
  Fujiwara, Ema, and Sasaki}}]{izumi16}
\bibinfo{author}{\bibfnamefont{S.}~\bibnamefont{Izumi}},
  \bibinfo{author}{\bibfnamefont{M.}~\bibnamefont{Takeoka}},
  \bibinfo{author}{\bibfnamefont{K.}~\bibnamefont{Wakui}},
  \bibinfo{author}{\bibfnamefont{M.}~\bibnamefont{Fujiwara}},
  \bibinfo{author}{\bibfnamefont{K.}~\bibnamefont{Ema}}, \bibnamefont{and}
  \bibinfo{author}{\bibfnamefont{M.}~\bibnamefont{Sasaki}},
  \bibinfo{journal}{Phys. Rev. A} \textbf{\bibinfo{volume}{94}},
  \bibinfo{pages}{033842} (\bibinfo{year}{2016}).

\bibitem[{\citenamefont{Paris}(1996)}]{paris96}
\bibinfo{author}{\bibfnamefont{M.~G.} \bibnamefont{Paris}},
  \bibinfo{journal}{Physics Letters A} \textbf{\bibinfo{volume}{217}},
  \bibinfo{pages}{78 } (\bibinfo{year}{1996}).

\bibitem[{\citenamefont{Gentile et~al.}(1996)\citenamefont{Gentile, Houston,
  and Cromer}}]{gentile96}
\bibinfo{author}{\bibfnamefont{T.~R.} \bibnamefont{Gentile}},
  \bibinfo{author}{\bibfnamefont{J.~M.} \bibnamefont{Houston}},
  \bibnamefont{and} \bibinfo{author}{\bibfnamefont{C.~L.}
  \bibnamefont{Cromer}}, \bibinfo{journal}{Appl. Opt.}
  \textbf{\bibinfo{volume}{35}}, \bibinfo{pages}{4392} (\bibinfo{year}{1996}).

\bibitem[{\citenamefont{Kikuchi and Tsukamoto}(2008)}]{kikuchi08}
\bibinfo{author}{\bibfnamefont{K.}~\bibnamefont{Kikuchi}} \bibnamefont{and}
  \bibinfo{author}{\bibfnamefont{S.}~\bibnamefont{Tsukamoto}},
  \bibinfo{journal}{Lightwave Technology, Journal of}
  \textbf{\bibinfo{volume}{26}}, \bibinfo{pages}{1817} (\bibinfo{year}{2008}).

\bibitem[{\citenamefont{{Xie} et~al.}(2011)\citenamefont{{Xie}, {Dang}, and
  {Guo}}}]{xie11}
\bibinfo{author}{\bibfnamefont{G.}~\bibnamefont{{Xie}}},
  \bibinfo{author}{\bibfnamefont{A.}~\bibnamefont{{Dang}}}, \bibnamefont{and}
  \bibinfo{author}{\bibfnamefont{H.}~\bibnamefont{{Guo}}}, in
  \emph{\bibinfo{booktitle}{2011 IEEE International Conference on Communications (ICC), Kyoto}}
   (\bibinfo{year}{2011}), pp. \bibinfo{pages}{1--6}.

\bibitem[{\citenamefont{{Ghozlan} and {Kramer}}(2013)}]{ghozlan13}
\bibinfo{author}{\bibfnamefont{H.}~\bibnamefont{{Ghozlan}}} \bibnamefont{and}
  \bibinfo{author}{\bibfnamefont{G.}~\bibnamefont{{Kramer}}}, in
  \emph{\bibinfo{booktitle}{2013 IEEE International Symposium on Information Theory}}
   (\bibinfo{year}{2013}), pp. \bibinfo{pages}{2279--2283}.

\bibitem[{\citenamefont{Khanzadi et~al.}(2015)}]{khanzadi15}
\bibinfo{author}{\bibfnamefont{M.}~\bibnamefont{Khanzadi}},
  \emph{\bibinfo{title}{Phase Noise in Communication Systems: Modeling, Compensation, and Performance Analysis, PhD Thesis}},
  \bibinfo{publisher}{Chalmers University of Technology, Gothenberg, Sweden},
  \bibinfo{year}{2015}.
  
\bibitem[{\citenamefont{Goldfarb and Li}(2006)}]{goldfarb06}
\bibinfo{author}{\bibfnamefont{G.}~\bibnamefont{Goldfarb}} \bibnamefont{and}
  \bibinfo{author}{\bibfnamefont{G.}~\bibnamefont{Li}}, \bibinfo{journal}{Opt.
  Express} \textbf{\bibinfo{volume}{14}}, \bibinfo{pages}{8043}
  (\bibinfo{year}{2006}).

\bibitem[{\citenamefont{{Salz}}(1986)}]{salz86}
\bibinfo{author}{\bibfnamefont{J.}~\bibnamefont{{Salz}}},
  \bibinfo{journal}{IEEE Communications Magazine}
  \textbf{\bibinfo{volume}{24}}, \bibinfo{pages}{38} (\bibinfo{year}{1986}).

\bibitem[{\citenamefont{Holzman and Ivry}(2019)}]{holzman2019}
\bibinfo{author}{\bibfnamefont{I.}~\bibnamefont{Holzman}} \bibnamefont{and}
  \bibinfo{author}{\bibfnamefont{Y.}~\bibnamefont{Ivry}},
  \bibinfo{journal}{Advanced Quantum Technologies}
  \textbf{\bibinfo{volume}{2}}, \bibinfo{pages}{1800058}
  (\bibinfo{year}{2019}).

\bibitem[{\citenamefont{Guan et~al.}(2016)\citenamefont{Guan, Xu, Yin, Li,
  Zhang, Chen, Yang, Li, You, Chen et~al.}}]{guan16}
\bibinfo{author}{\bibfnamefont{J.-Y.} \bibnamefont{Guan}},
  \bibinfo{author}{\bibfnamefont{F.}~\bibnamefont{Xu}},
  \bibinfo{author}{\bibfnamefont{H.-L.} \bibnamefont{Yin}},
  \bibinfo{author}{\bibfnamefont{Y.}~\bibnamefont{Li}},
  \bibinfo{author}{\bibfnamefont{W.-J.} \bibnamefont{Zhang}},
  \bibinfo{author}{\bibfnamefont{S.-J.} \bibnamefont{Chen}},
  \bibinfo{author}{\bibfnamefont{X.-Y.} \bibnamefont{Yang}},
  \bibinfo{author}{\bibfnamefont{L.}~\bibnamefont{Li}},
  \bibinfo{author}{\bibfnamefont{L.-X.} \bibnamefont{You}},
  \bibinfo{author}{\bibfnamefont{T.-Y.} \bibnamefont{Chen}},
  \bibnamefont{et~al.}, \bibinfo{journal}{Phys. Rev. Lett.}
  \textbf{\bibinfo{volume}{116}}, \bibinfo{pages}{240502}
  (\bibinfo{year}{2016}).

\bibitem[{\citenamefont{Takeoka et~al.}(2008)\citenamefont{Takeoka, Sasaki}}]{takeoka08}
\bibinfo{author}{\bibfnamefont{M.}~\bibnamefont{Takeoka}}, \bibnamefont{and}
  \bibinfo{author}{\bibfnamefont{M.} \bibnamefont{Sasaki}},
  \bibinfo{journal}{Phys. Rev. A} \textbf{\bibinfo{volume}{78}},
  \bibinfo{pages}{022320} (\bibinfo{year}{2008}).

\bibitem[{\citenamefont{{Proakis} and {Salehi}}(2000)}]{proakis00}
\bibinfo{author}{\bibfnamefont{J.}~\bibnamefont{{Proakis}}} \bibnamefont{and}
  \bibinfo{author}{\bibfnamefont{M.}~\bibnamefont{{Salehi}}}, in
  \emph{\bibinfo{booktitle}{Digital Communication, 5th Edition}} (\bibinfo{publisher}{McGraw-Hill, New York}, \bibinfo{year}{2000}) .

  \end{thebibliography}
\end{document}